\documentclass[journal]{IEEEtran}
\usepackage{amsmath,amssymb,amsfonts}
\usepackage{float}
\usepackage{multirow}
\usepackage{graphicx}
\usepackage{tabularx}

\usepackage{longtable}  % Allows long tables to span multiple pages
\usepackage{graphicx}   % For better table formatting
\usepackage{booktabs}   % For more advanced table formatting

\graphicspath{%
    {converted_graphics/}% inserted by PCTeX
    {/}% inserted by PCTeX
}
  \usepackage{xcolor}
  \makeatletter
\newcommand*{\rom}[1]{\expandafter\@slowromancap\romannumeral #1@}
\makeatother

\usepackage{makecell}
\usepackage{cite}
\ifCLASSINFOpdf
\else
\fi
\begin{document}
 % \bstctlcite{IEEEexample:BSTcontrol}
\title{ML-Enabled Open RAN: A Comprehensive  Survey of Architectures, Challenges, and Opportunities}

 \author{Mira Chandra Kirana,~\IEEEmembership{Student Member, IEEE}, Patatchona Keyela,~\IEEEmembership{Student Member, IEEE}, Fatemeh Rostamian,~\IEEEmembership{Student Member, IEEE}, Deemah H. Tashman,~\IEEEmembership{Member, IEEE} and  Soumaya Cherkaoui,~\IEEEmembership{Senior Member, IEEE}
\thanks{The Department of Computer and Software Engineering, Polytechnique Montreal, Montreal, QC, Canada,  H3T 1J4 (e-mail: \{mira-chandra.kirana,keyela.patatchona,fatemeh.rostamian,deemah.tashman, soumaya.cherkaoui\}@polymtl.ca).} }
\maketitle


\begin{abstract}
As wireless communication systems become more advanced, Open Radio Access Networks (O-RAN) stand out as a notable framework that promotes interoperability and cost-effectiveness. An examination of the progression of RAN architectures, as well as O-RAN’s underlying principles, reveals the importance of machine learning (ML) in addressing various challenges, including spectrum management, resource allocation, and security. Hence, this survey provides a comprehensive overview of the integration of ML within O-RAN, highlighting its transformative potential in enhancing network performance and efficiency. This survey aims to describe the current status of ML applications in O-RAN while indicating possible directions for future research by analyzing existing literature. The findings aim to assist researchers and stakeholders in formulating optimal service strategies and advancing the understanding of intelligent wireless networks.
\end{abstract}


% \fi
% \IEEEpeerreviewmaketitle

\section{Introduction}

 \IEEEPARstart{S}{ince} the introduction of the open radio access network (O-RAN) concept in 2018, there has been growing academic and industrial interest in applying machine learning (ML) to enhance its functionality. Although early research on ML in O-RAN was limited, the field gained traction starting with a seminal 2020 paper that outlined the evolution of RAN architectures and introduced the foundational concepts of O-RAN as a next-generation solution \cite{Singh_The_Evolution_2020}. By the end of 2020, several studies emerged exploring ML's role in optimizing O-RAN performance, marking the beginning of a rapidly expanding research area.
 
O-RAN is designed to disaggregate traditional RAN components, thereby promoting interoperability, cost efficiency, and innovation through open interfaces and virtualization. To fully realize these benefits, ML is a key enabler, where ML techniques can enhance the intelligence and flexibility of O-RAN by addressing challenges such as resource allocation, mobility management, anomaly detection, and dynamic traffic optimization \cite{li_rlops_2022,lee_o-ran_2021,giannopoulos_supporting_2022}. These capabilities are largely facilitated through architectural elements like the Non-Real-Time (Non-RT) and the Near-Real-Time (Near-RT) RAN Intelligent Controllers (RICs), which support policy-driven ML integration \cite{Ndikumana_Federated_2023}.

In broader 5G and 6G contexts, ML in O-RAN plays a pivotal role in addressing critical issues such as efficient resource management, service quality optimization, and network security \cite{haryo_prananto_o-ran_2022,abedin_elastic_2022}. It enables fine-grained modeling and control of network resources, ranging from power distribution and routing to traffic and interference management \cite{lee_hosting_2020,iturria-rivera_multi-agent_2022,ramezanpour_intelligent_2022}, and ultimately leads to more intelligent and adaptive wireless systems \cite{perveen_dynamic_2023,Sen_Intelligent_2022,Nagib_Safe_2024}.

In O-RAN contexts, reinforcement learning (RL) is widely used due to its capacity to optimize resource allocation, improve network performance, and effectively respond to dynamic network conditions.  RL algorithms facilitate the acquisition of knowledge by O-RAN systems through interactions with the environment, allowing them to make decisions that prioritize maximizing rewards or accomplishing specified objectives \cite{Sharara_Reinforcement_2023,Cheng_Reinforcement_2023,houda_federated_2024}. An essential benefit of RL in O-RAN is its capacity to manage intricate and ever-changing network settings effectively. RL models have the ability to adjust to various network conditions, including fluctuating traffic loads, user requirements, and degrees of interference, through ongoing learning and updating of their decision-making policies \cite{kumar_quality_2023,Amiri_DeepReinforcement_2024,Vila_Training_2022}. In the O-RAN scenario, adaptability is crucial due to rapid fluctuations in network conditions, which require instant optimization and decision-making. 
Thanks to its advantages, including flexibility, adaptability to environmental changes, continuous optimization, and the ability to overcome complex security challenges, RL has become the preferred method for ML integration in O-RAN \cite{shi_how_2023}. Nevertheless, this does not imply that supervised learning (SL) and unsupervised learning (UL) have little chance of advancing their integration with O-RAN, which presents several challenges and opportunities for further investigation.

 O-RAN offers flexibility and cost-efficiency, representing a transformative shift in cellular network architecture. However, these advantages come with multiple challenges that must be addressed, including complex supply chains, data confidentiality, and the seamless integration of artificial intelligence (AI) technologies within an open, multi-vendor, cloud-based environment. For instance, due to the dynamic nature of O-RAN and the heterogeneous deployment of network elements, spectrum management is becoming more complicated. To guarantee efficient and real-time spectrum allocation while avoiding interference and achieving fairness among users requires intelligent and adaptive solutions. Moreover, due to the disaggregation and virtualization of the O-RAN architecture, resource allocation is challenging, which means that coordinating resources across multi-vendor components requires intelligent orchestration, where ML, particularly RL, offers dynamic, real-time solutions.  However, network heterogeneity, latency, and scalability remain key obstacles to reliable and efficient deployment. Furthermore, the open nature and the multi-vendor integration of O-RAN can make networks more vulnerable to cyberattacks and data breaches, necessitating intelligent and adaptable security measures  \cite{liyanage2023open}. 
 In this survey, we delve into these key challenges through the lens of ML explore how ML techniques can be leveraged to develop intelligent, adaptive, and secure solutions tailored to the unique characteristics of O-RAN environments.


{\color{black}
\begin{table*}[ht]
\centering
\caption{List of  acronyms and definitions}
\begin{tabular}{|l|l|l|l|}
    \hline
    \textbf{Acronym} & \textbf{Definition} & \textbf{Acronym} & \textbf{Definition} \\
    \hline
    3GPP & 3rd Generation Partnership Project & A2C & Advantage Actor-Critic \\
    \hline
    ACER & Actor-Critic with Experience Replay & AI & Artificial Intelligence \\
    \hline
    API & Application Programming Interfaces & ARIMA & AutoRegressive Integrated Moving Average \\
    \hline
    BBU & Baseband Unit & BS & Base Station \\
    \hline
    CAPEX & Capital Expenditure & CNN & Convolutional Neural Network \\
    \hline
    COTS & Commercial Off-The-Shelf & CPRI & Common Public Radio Interface \\
    \hline
    CR & Cognitive Radio & CRAN & Cloud Radio Access Network \\
    \hline
    CTI & Cyber Threat Intelligence & CU & Centralized Unit \\
    \hline
    DL & Deep Learning & DQN & Deep Q-Network \\
    \hline
    DRL & Deep Reinforcement Learning & DSA & Dynamic Spectrum Access \\
    \hline
    F-DQN & Federated Deep Q-Network & F-DRL & Federated Deep Reinforcement Learning \\
    \hline
    F-MARL & Federated Multi-Agent Reinforcement Learning & FedAvg & Federated Averaging \\
    \hline
    FFNN & Feedforward Neural Network & FL & Federated Learning \\
    \hline
    FRL & Federated Reinforcement Learning & GBT & Gradient Boosted Trees \\
    \hline
    GNB & Gaussian Naïve Bayes & HARQ & Hybrid Automatic Repeat Request \\
    \hline
    HRL & Hierarchical Reinforcement Learning & IDS & Intrusion Detection System \\
    \hline
    IF & Isolation Forest & IoT & Internet of Things \\
    \hline
    K-Means & K-Means Clustering & {\color{black} KPMs} & Key Performance measurements \\
    \hline
    KNN & K-Nearest Neighbors & LSTM & Long Short-Term Memory \\
    \hline
    MAB & Multi-Armed Bandit & MAC & Medium Access Control \\
    \hline
    MADRL & Multi-Agent Deep Reinforcement Learning & MARL & Multi-Agent Reinforcement Learning \\
    \hline
    MCS & Modulation and Coding Scheme & MCTS & Monte Carlo Tree Search \\
    \hline
    MEC & Mobile Edge Computing & ML & Machine Learning \\
    \hline
    mMTC & Massive Machine-Type Communication & MultiRATs & Multiple Radio Access Technologies \\
    \hline
    {\color{black} NIB} & Network information base & NN & Neural Network \\
    \hline
    NR & New Radio (5G) & {\color{black} OAM} & Services Operations, administration, and maintenance \\
    \hline
    ONAP & Open Network Automation Platform & OpenFM & Open Fault Management \\
    \hline
    OpEx & {\color{black} Operating Expenditure} & O-RAN & Open Radio Access Network \\
    \hline
    {\color{black} PDCP} & Packet Data Convergence Protocol & PPO & Proximal Policy Optimization \\
    \hline
    PRB & Physical Resource Block & PUE & Primary User Emulation \\
    \hline
    Q-Learning & Q-Learning (Reinforcement Learning Algorithm) & QoE & Quality of Experience \\
    \hline
    QoS & Quality of Service & RAN & Radio Access Network \\
    \hline
    RB & Resource Block & RBG & Resource Block Group \\
    \hline
    RF & Random Forest / Radio Frequency (context-dependent) & RIC & RAN Intelligent Controller \\
    \hline
    RL & Reinforcement Learning & RNN & Recurrent Neural Network \\
    \hline
    RRH & Remote Radio Head & RRM & Radio Resource Management \\
    \hline
    {\color{black} RRC} & Radio Resource Control & RU & Radio Unit \\
    \hline
    SAGINs & Space-Air-Ground Integrated Networks & SARSA & State-Action-Reward-State-Action Algorithm \\
    \hline
    {\color{black} SDAP} & Service Data Adaptation Protocol & SDN & Software-Defined Networking \\
    \hline
    {\color{black} SDL} & Shared data Layer & SLA & Service Level Agreement \\
    \hline
    SL & Supervised Learning & SMO & Service Management and Orchestration \\
    \hline
    {\color{black} SON} & Self-Organizing Networks & SS & Spectrum sharing \\
    \hline
    SSDF & Spectrum Sensing Data Falsification & SVM & Support Vector Machine \\
    \hline
    UAV & Unmanned Aerial Vehicle & UE & User Equipment \\
    \hline
    {\color{black} UP} & User plane & URLLC & Ultra-Reliable Low Latency Communication \\
    \hline
    VBS & Virtual Base Station & VFs & Virtual Functions \\
    \hline
    ViT & Vision Transformer & VNF & Virtual Network Functions \\
    \hline
    VRAN & Virtual Radio Access Network & WG2 & Working Group 2 (of O-RAN Alliance) \\
    \hline
    XAI & Explainable AI & ZTA & Zero Trust Architecture \\
    \hline
\end{tabular}
\label{tab:acronyms}
\end{table*}
}

{\color{black}
\textbf{Related works:}
% Related works (only surveys) and Comparison between our survey and previous ones
The literature has extensively investigated the integration of ML into O-RAN, with numerous essential studies providing valuable insights.  Many current studies focus on specific aspects of ML, for instance, some on Deep Learning (DL) and its ability to enhance the functionality of Self-Organizing Networks (SONs) within the O-RAN framework \cite{Brik_DeepLearning_2022}.  Others are more case-specific, providing summaries of how data-driven, autonomous, and self-optimizing ML capabilities can enable resource management for RAN slicing in 5G and beyond \cite{azimi_applications_2022, bartsiokas_ml-based_2022}.  Further research has underscored the security challenges ML integration in O-RAN introduces, emphasizing the vulnerabilities that could be exploited \cite{chen2023brief, amachaghi_survey_2024}. At the same time, AI/ML can also be a solution, particularly for improving O-RAN security through anomaly-detection and attack-detection techniques \cite{you_survey_2023}, with Smart IoT showing great potential as an intelligent use case for security-aware O-RAN applications \cite{musa_open_2023}.
Network automation is also one of the important benefits of AI/ML integration as an intelligent component in the O-RAN architecture \cite{hamdan_recent_2023}. Likewise, energy consumption optimization is a key focus in the existing literature, given that ML training and inference processes require resources \cite{liang_energy_2024}. The development of research on the integration and utilization of AI/ML in O-RAN is inseparable from the needs and challenges of datasets that match the tasks of the AI/ML models to be developed, which encouraged the authors in \cite{couto_survey_2024} to provide survey results regarding the availability of datasets in O-RAN.

Table \ref{tab:survey_ml_in_O-RAN} summarizes the surveys conducted in the domain of ML in O-RAN.  
Most surveys focused on specific challenges, such as resource allocation, energy consumption, network automation, and O-RAN security, separately. In addition, only two papers discussed all types of ML approaches in O-RAN. However, the challenges addressed are limited to Radio Resource Management (RRM) and energy consumption, with contributions focused on modeling RRM to improve efficiency and AI/ML procedures in energy-intensive O-RAN architectures. After all, the increasing demand for wireless communication is directly proportional to the rising challenges in all aspects of O-RAN, motivating us to explore the utilization and integration of ML in O-RAN more deeply.  The table reveals that many papers lack a comprehensive treatment of O-RAN’s evolution, architectural design, the application of diverse AI/ML techniques to key challenges, and strategic insights into future research directions for ML in O-RAN. Hence, after reviewing the existing work, we highlight that our work offers a deeper and broader approach, which aligns with the growing O-RAN challenges. Dissimilar with others, our work not only focuses on specific problems but also provides a more thorough understanding of the use of all types of ML in O-RAN, addresses challenges in spectrum management, resource allocation, and security, and provides more applicable and needed research directions.

\begin{table*}[!t]
\caption{Surveys of AI/ML in O-RAN}
\label{tab:survey_ml_in_O-RAN}
\centering
\begin{tabular}{|p{1.3cm}|p{0.4cm}|p{0.4cm}|p{0.4cm}|p{0.4cm}|p{1.0cm}|p{1.3cm}|p{1.8cm}|p{6.0cm}|}
\hline
\multicolumn{1}{|c|}{\multirow{3}{*}{\textbf{Ref/Year}}} & \multirow{3}{*}{\textbf{SL}} & \multirow{3}{*}{\textbf{UL}} & \multirow{3}{*}{\textbf{RL}} & \multirow{3}{*}{\textbf{FL}} & \multicolumn{3}{c|}{\textbf{The Challenges Tackled by ML Approach}} & \multicolumn{1}{c|}{\multirow{3}{*}{\textbf{Contributions Summary}}}\\
\cline{6-8} 
& & & & & \multirow{2}{*}{\textbf{Security}} & \textbf{Resource Allocation} & \textbf{Spectrum Management} & \multicolumn{1}{c|}{} \\
\hline 

\multicolumn{1}{|c|}{\multirow{4}{*}{\cite{Brik_DeepLearning_2022}/2022}} & \multicolumn{1}{c|}{\multirow{4}{*}{\checkmark}} & \multicolumn{1}{c|}{\multirow{4}{*}{-}} & \multicolumn{1}{c|}{\multirow{4}{*}{\checkmark}} & \multicolumn{1}{c|}{\multirow{4}{*}{-}} & \multicolumn{1}{c|}{\multirow{4}{*}{-}} & \multicolumn{1}{c|}{\multirow{4}{*}{\checkmark}} & \multicolumn{1}{c|}{\multirow{4}{*}{\checkmark}} &  {\color{black} Provides a thorough review of the application of DL to O-RAN architecture through case studies and demonstrates consistent performance by automating DL modeling.}  \\
\hline

\multicolumn{1}{|c|}{\multirow{7}{*}{\cite{azimi_applications_2022}/2022}} & \multicolumn{1}{c|}{\multirow{7}{*}{\checkmark}} & \multicolumn{1}{c|}{\multirow{7}{*}{\checkmark}} & \multicolumn{1}{c|}{\multirow{7}{*}{\checkmark}} & \multicolumn{1}{c|}{\multirow{7}{*}{\checkmark}} & \multicolumn{1}{c|}{\multirow{7}{*}{-}} & \multicolumn{1}{c|}{\multirow{7}{*}{\checkmark}} & \multicolumn{1}{c|}{\multirow{7}{*}{-}} & {\color{black} Classifies ML techniques used in resource slicing management, analyze each study based on the algorithms used, challenges overcome, and types of resources allocated, compare various methods based on performance and efficiency parameters in RAN slicing, and identify practical challenges and future research directions.}  \\
\hline

\multicolumn{1}{|c|}{\multirow{3}{*}{\cite{bartsiokas_ml-based_2022}/2022}} & \multicolumn{1}{c|}{\multirow{3}{*}{\checkmark}} & \multicolumn{1}{c|}{\multirow{3}{*}{\checkmark}} & \multicolumn{1}{c|}{\multirow{3}{*}{\checkmark}} & \multicolumn{1}{c|}{\multirow{3}{*}{-}} & \multicolumn{1}{c|}{\multirow{3}{*}{-}} & \multicolumn{1}{c|}{\multirow{3}{*}{\checkmark}} & \multicolumn{1}{c|}{\multirow{3}{*}{-}} & {\color{black} Develops research framework guidelines for the efficient management of resources in 5G and beyond through the use of AI/ML.}  \\
\hline

\multicolumn{1}{|c|}{\multirow{2}{*}{\cite{chen2023brief}/2023}} & \multicolumn{1}{c|}{\multirow{2}{*}{-}} & \multicolumn{1}{c|}{\multirow{2}{*}{-}} & \multicolumn{1}{c|}{\multirow{2}{*}{-}} & \multicolumn{1}{c|}{\multirow{2}{*}{-}} & \multicolumn{1}{c|}{\multirow{2}{*}{\checkmark}} & \multicolumn{1}{c|}{\multirow{2}{*}{-}} & \multicolumn{1}{c|}{\multirow{2}{*}{-}} & {\color{black}  Integrates of AI/ML is one of the components of the O-RAN that are susceptible to attack.} \\
\hline

\multicolumn{1}{|c|}{\multirow{6}{*}{\cite{amachaghi_survey_2024}/2024}} & \multicolumn{1}{c|}{\multirow{6}{*}{-}} & \multicolumn{1}{c|}{\multirow{6}{*}{-}} & \multicolumn{1}{c|}{\multirow{6}{*}{\checkmark}} & \multicolumn{1}{c|}{\multirow{6}{*}{\checkmark}} & \multicolumn{1}{c|}{\multirow{6}{*}{\checkmark}} & \multicolumn{1}{c|}{\multirow{6}{*}{-}} & \multicolumn{1}{c|}{\multirow{6}{*}{-}} & {\color{black} Provides a survey of the security aspects of O-RAN, introduce a structured taxonomy of O-RAN security threats, provide an in-depth analysis of Intrusion Detection System (IDS) in O-RAN environments, and provide a case study describing security integration in O-RAN deployments.}  \\
\hline

\multicolumn{1}{|c|}{\multirow{3}{*}{\cite{you_survey_2023}/2023}} & \multicolumn{1}{c|}{\multirow{3}{*}{-}} & \multicolumn{1}{c|}{\multirow{3}{*}{-}} & \multicolumn{1}{c|}{\multirow{3}{*}{-}} & \multicolumn{1}{c|}{\multirow{3}{*}{-}} & \multicolumn{1}{c|}{\multirow{3}{*}{\checkmark}} & \multicolumn{1}{c|}{\multirow{3}{*}{-}} & \multicolumn{1}{c|}{\multirow{3}{*}{-}} & {\color{black} Reviews the security issues and solutions in the space-air-ground integrated network (SAGIN) 6G, particularly threats to AI-enabled O-RAN.}  \\
\hline

\multicolumn{1}{|c|}{\multirow{4}{*}{\cite{musa_open_2023}/2023}} & \multicolumn{1}{c|}{\multirow{4}{*}{\checkmark}} & \multicolumn{1}{c|}{\multirow{4}{*}{-}} & \multicolumn{1}{c|}{\multirow{4}{*}{\checkmark}} & \multicolumn{1}{c|}{\multirow{4}{*}{-}} & \multicolumn{1}{c|}{\multirow{4}{*}{\checkmark}} & \multicolumn{1}{c|}{\multirow{4}{*}{-}} & \multicolumn{1}{c|}{\multirow{4}{*}{-}} & {\color{black} Provides a comprehensive examination of the implementation and dimensions of O-RAN issues in smart IoT, potential security hazards, and mitigation strategies.}  \\
\hline

\multicolumn{1}{|c|}{\multirow{5}{*}{\cite{hamdan_recent_2023}/2023}} & \multicolumn{1}{c|}{\multirow{5}{*}{\checkmark}} & \multicolumn{1}{c|}{\multirow{5}{*}{-}} & \multicolumn{1}{c|}{\multirow{5}{*}{\checkmark}} & \multicolumn{1}{c|}{\multirow{5}{*}{\checkmark}} & \multicolumn{1}{c|}{\multirow{5}{*}{\checkmark}} & \multicolumn{1}{c|}{\multirow{5}{*}{\checkmark}} & \multicolumn{1}{c|}{\multirow{5}{*}{-}} & {\color{black} An overview of O-RAN architecture and components, exploration of challenges in ML-based automation in O-RAN, application of ML algorithms in O-RAN, and research opportunities based on the benefits of ML in O-RAN are presented.}  \\
\hline

\multicolumn{1}{|c|}{\multirow{7}{*}{\cite{liang_energy_2024}/2024}} & \multicolumn{1}{c|}{\multirow{7}{*}{\checkmark}} & \multicolumn{1}{c|}{\multirow{7}{*}{\checkmark}} & \multicolumn{1}{c|}{\multirow{7}{*}{\checkmark}} & \multicolumn{1}{c|}{\multirow{7}{*}{\checkmark}} & \multicolumn{1}{c|}{\multirow{7}{*}{-}} & \multicolumn{1}{c|}{\multirow{7}{*}{\checkmark}} & \multicolumn{1}{c|}{\multirow{7}{*}{-}} &  {\color{black} Provides an explanation of the architectural components and open interfaces of O-RAN,   the background and recent ML methods in O-RAN, a comprehensive review of energy consumption during the training and inference phases of ML in O-RAN, and   a case study showing a real scenario for energy consumption in O-RAN.}  \\
\hline

\multicolumn{1}{|c|}{\multirow{4}{*}{\cite{couto_survey_2024}/2024}} & \multicolumn{1}{c|}{\multirow{4}{*}{-}} & \multicolumn{1}{c|}{\multirow{4}{*}{-}} & \multicolumn{1}{c|}{\multirow{4}{*}{-}} & \multicolumn{1}{c|}{\multirow{4}{*}{-}} & \multicolumn{1}{c|}{\multirow{4}{*}{-}} & \multicolumn{1}{c|}{\multirow{4}{*}{-}} & \multicolumn{1}{c|}{\multirow{4}{*}{-}} &  {\color{black} Identifies the significant O-RAN datasets, and provide classification cases using ChARM (Channel-Aware Resource Management) and Colosseum O-RAN COMMAG datasets.}  \\
\hline

\multicolumn{1}{|c|}{\multirow{5}{*}{This paper}} & \multicolumn{1}{c|}{\multirow{5}{*}{\checkmark}} & \multicolumn{1}{c|}{\multirow{5}{*}{\checkmark}} & \multicolumn{1}{c|}{\multirow{5}{*}{\checkmark}} & \multicolumn{1}{c|}{\multirow{5}{*}{\checkmark}} & \multicolumn{1}{c|}{\multirow{5}{*}{\checkmark}} & \multicolumn{1}{c|}{\multirow{5}{*}{\checkmark}} & \multicolumn{1}{c|}{\multirow{5}{*}{\checkmark}} & {\color{black} This survey offers a comprehensive overview of O-RAN and AI/ML integration, as well as strategic research directions that should be developed and adapted by all relevant stakeholders to address all challenges in O-RAN.}  \\
\hline

\end{tabular}
\end{table*}

{\color{black}
\textbf{Motivation:} Considering the aforementioned surveys on O-RAN and ML, no existing work provides a comprehensive review that simultaneously addresses spectrum management, resource allocation, and security as critical challenges in O-RAN using ML techniques. Most prior surveys focus on only one or two of these aspects, limiting a holistic understanding of their interplay within ML-enabled O-RAN systems. To address this gap, our paper presents an extensive review of ML applications in O-RAN to tackle these essential challenges, complemented by illustrative case studies, while also identifying areas that remain underexplored. Drawing on these insights, we outline targeted future research directions to guide the development of more adaptive, secure, and intelligent O-RAN networks capable of meeting the demands of next-generation wireless services.

%Although machine learning has become increasingly central to the evolution of O-RAN, existing surveys typically examine spectrum management, resource allocation, or security in isolation, offering only narrow insights into challenges that are, in reality, deeply interconnected. Treating these domains separately overlooks how decisions in one area can shape constraints, vulnerabilities, and opportunities in the others. To address this fragmentation, our survey provides an integrated examination of all three challenges, highlighting the ways in which ML techniques can operate across them to support more coherent and efficient O-RAN design. This unified perspective is strengthened through a set of representative case studies that demonstrate how ML-driven approaches can jointly influence spectrum usage, resource provisioning, and security posture in practical settings. Drawing on these insights, we further identify the open issues that arise from the interplay of these domains and outline targeted future research directions to guide the development of more adaptive, secure, and intelligent O-RAN systems.

\textbf{Contributions:} Following the motivation of the paper, the main contributions of this survey are given as follows:  

\begin{itemize}
    \item Identification of open challenges: We highlight key O-RAN challenges that can be addressed through ML approaches, specifically in spectrum management, resource allocation, and security, emphasizing the need for innovative solutions and extensive experimentation.
    
    \item Illustrative case studies: Two case studies demonstrate the practical impact of ML: deep reinforcement learning (DRL) for resource allocation and SL for security, showing how ML techniques can enhance critical O-RAN functionalities.

    \item Structured ML taxonomy: We present a concise taxonomy that organizes AI/ML usage in O-RAN across three primary objectives—service quality enhancement, communication quality enhancement, and security quality enhancement. Each objective is linked to its core challenges and associated ML techniques. We also summarize the main advantages and limitations of ML in O-RAN, offering a balanced perspective on its potential and practical constraints.

    \item Future research directions: We outline promising avenues for advancing O-RAN, including conflict mitigation in multi-component systems, integration of millimeter-wave and terahertz technologies, scalability and performance optimization, adoption of ultra-massive MIMO, efficiency improvements via mobile edge computing (MEC), and leveraging digital twins to support stringent URLLC requirements.
\end{itemize}
}

\textbf{Structure of the Survey:} As shown in Fig.  \ref{fig:pict_structure}, the subsequent sections of the survey are organized as follows: Section II provides an overview of O-RAN and its underlying architecture, including the foundational principles of O-RAN, the evolution of RAN architectures, and the roles of the primary architectural components. In Section III, we examine the use of ML techniques in O-RAN, accompanied by an extensive literature review of research in this field, the advantages and practical limitations of ML in O-RAN, and the taxonomy of ML techniques in O-RAN context. Section IV outlines ML's capabilities for addressing O-RAN key challenges in spectrum management, resource allocation, and security, supported by representative case studies. Section V highlights open research directions for applying ML in O-RAN, including conflict mitigation, mmWave and Terahertz integration, scalability and performance optimization in large-scale O-RAN, ultra-massive MIMO, MEC integration with O-RAN, and digital twin technology, to encourage further investigation in these areas.  Finally, Section VI concludes the survey with a summary of key insights.}

\begin{figure}[h!]
\centering
\includegraphics[width=3.5in]{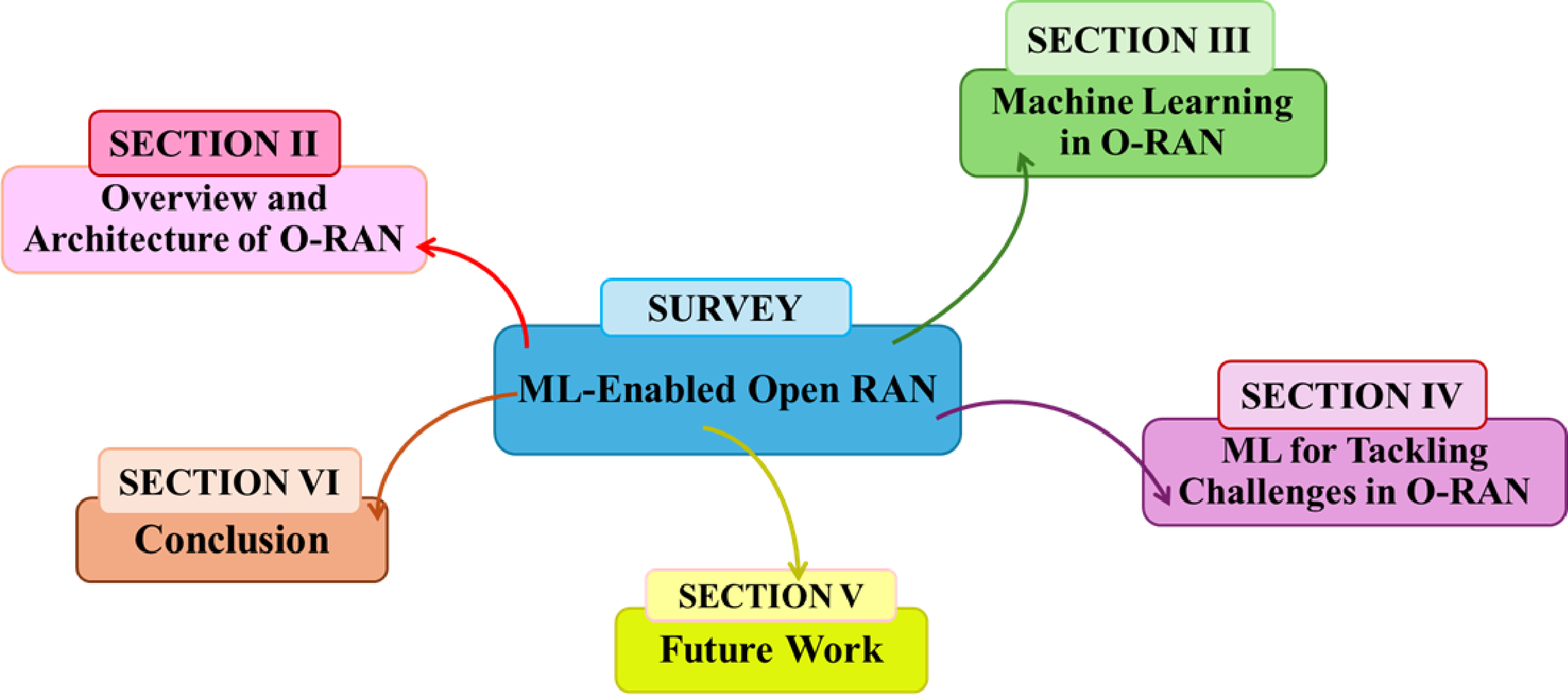}
\caption{Structure of Survey.}
\label{fig:pict_structure}
\end{figure}

{\color{black}
\section{Overview and Evolution of O-RAN}
\label{sec:openran}
Wireless mobile communication systems have been continuously evolving with various quality of service (QoS) \cite{8579399} requirements to enable different innovative applications such as IoT systems \cite{patatchona_csma2022,keyela2022analytical}, autonomous vehicles, and smart cities \cite{10.4108/eetiot.v9i3.2213, 10.61925/swb.2023.1303}. The RAN has been a central part of this evolution as the critical link facilitating communication between mobile devices and the core network infrastructure. RAN efficiency highly determines the data throughput, network coverage, user experience, and network operational efficiency and flexibility \cite{Saily.2020}. Traditional RAN architectures, such as distributed RAN (D-RAN) and cloud RAN (C-RAN), offered proximity advantages with lower latency,  direct connections between the radio units and the baseband units (BBU), and centralized processing for improved resource allocation \cite{2019.2899663, 2023.10338839, ett.3778, 2018.2865955, 2015.7143329, 2021.3104660}. However, scalability challenges and vendor dependency became limiting factors for innovation and adaptability in the network deployment and management. The hardware abstraction in virtual RAN (vRAN) with the decoupling of network functions (NF) improved the flexibility and cost efficiency, but the complexities and demands in recent network generations like 5G and beyond, revealed the need for modular, open, and interoperable RANs \cite{Khan.2022,9415591,ahmad2020,Ma.2020}.

This section provides the evolution towards the O-RAN framework, its comprehensive overview, foundational principles, main architectural elements, and the operational benefits it brings within the telecommunications domain.

\subsection{Evolution of RAN Architecture}
\label{subsec:RAN-evolution}
The RAN in mobile wireless networks, as shown in Fig. \ref{fig:dran}, connects the user equipment to the Core Network (CN) through the air interface. From the first generation (1G) to the latest 5G networks, RAN evolution has remarkably transformed the field of mobile telecommunications. An overview of how RAN has evolved over the years, along with the various factors that have led to this transformation, is presented in the following subsections.

\begin{figure}[!htbp]
\centering
\includegraphics[width=0.95\linewidth]{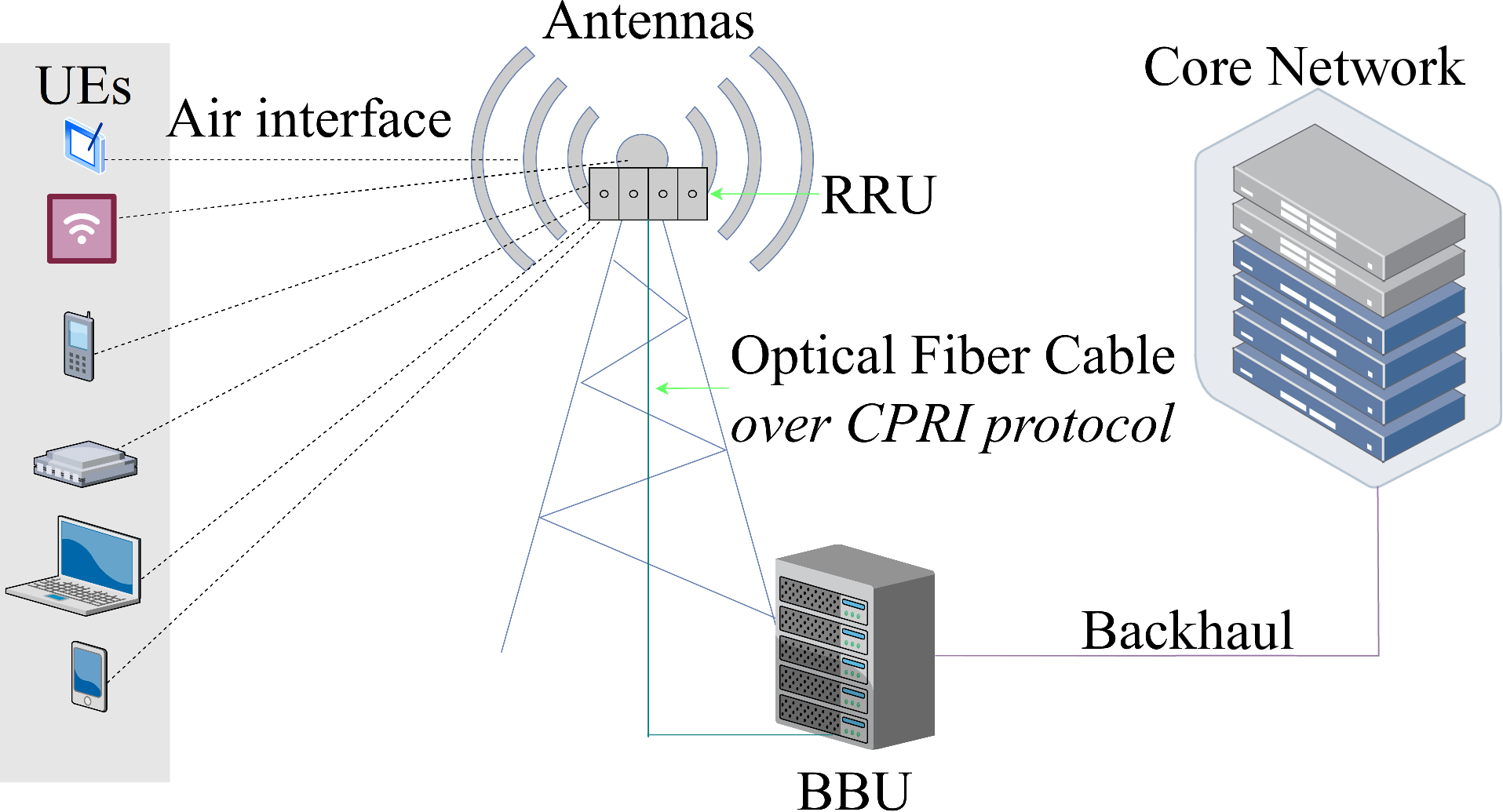}
\caption{A typical RAN serving different UEs}
\label{fig:dran}
\end{figure}

\subsubsection{Distributed RAN}
A RAN in 1G networks consisted of antennas and  a base station (BS), which was  a set of two elements: the radio unit (RU) and the BBU. The antennas were connected to RU and BBU through the radio frequency (RF) cabling. This version quickly evolved into having the RU remotely installed closer to antennas on the tower or in an elevated place, as illustrated in Fig. \ref{fig:dran} with the remote RU (RRU) interfacing the BBU through a fiber cable over the proprietary common public radio interface (CPRI) protocol \cite{8723481}, known as the backhaul interface. This setup was referred to as distributed RAN (D-RAN) since every RRU was served by its own BBU located in a secured room on the BS site, and all BBUs are directly connected to the CN through the backhaul interface, as illustrated in Fig. \ref{fig:dran-all}.

\begin{figure}[!htbp]
\centering
\includegraphics[width=0.9\linewidth]{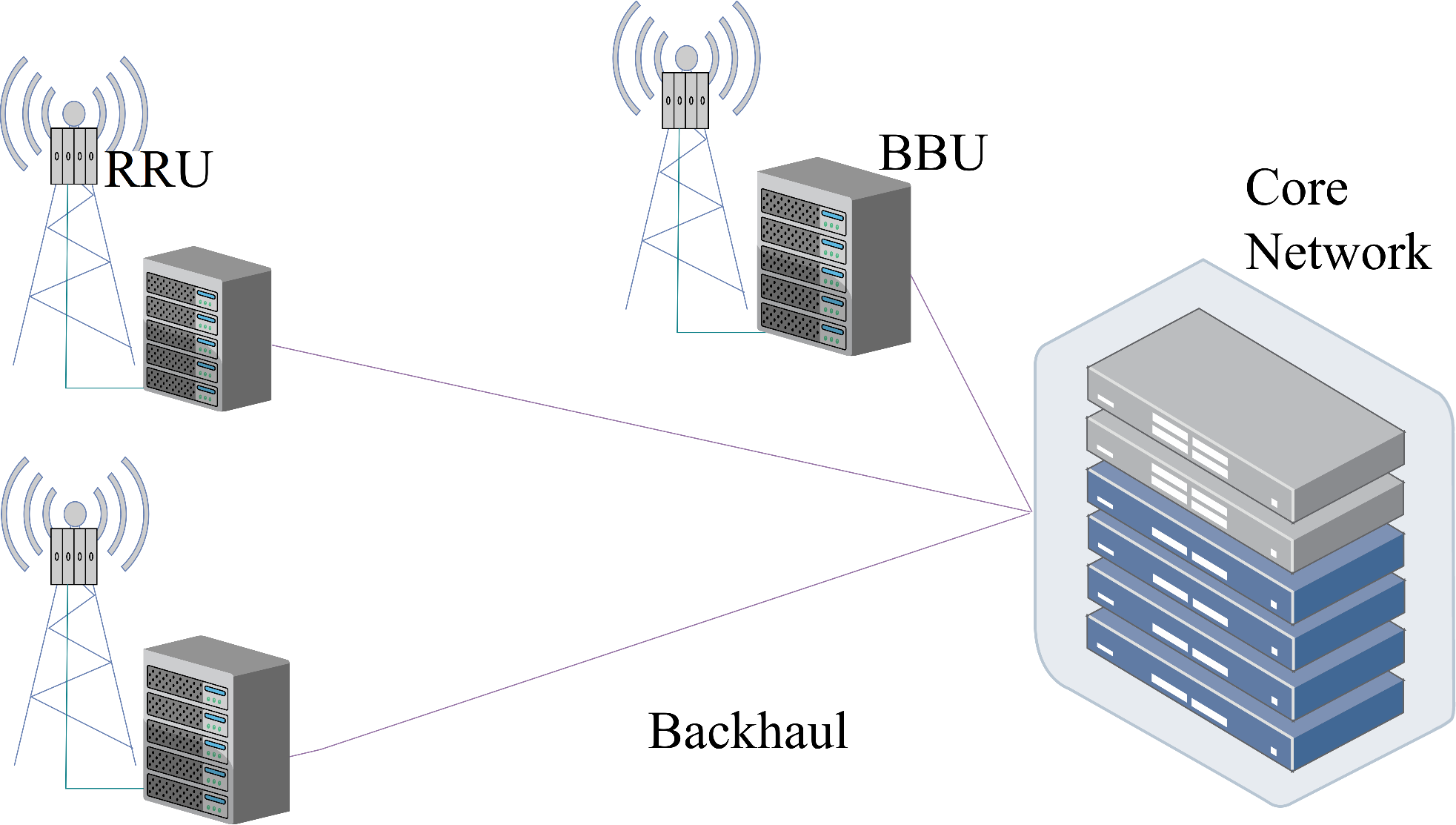}
\caption{Architecture of Distributed RAN}
\label{fig:dran-all}
\end{figure}

D-RANs had a  straightforward and rigid configuration, which facilitates communication between mobile devices and the network's core infrastructure, and were easy to deploy.  The proximity between the components reduces the need for complex, high-speed interfaces, as the transmission distances are minimal. Since each BS functions independently, D-RANs had a stable and consistent network performance with no reliance on centralized resources. 

However D-RANs were limited when the demand of higher data rates, connectivity, and the advent of new services kept growing, and the network operators faced the challenge of making their networks denser and requiring the deployment of additional BS, which led to a significant escalation in CapEx and operating expenditure (OpEx), including land leasing for BS, power consumption and cooling systems\cite{2019.2899663, 2023.10338839, 8723481, 6934902}. Moreover, the hardware and software of the RRU and the BBU, and the CPRI protocol that connects them, are proprietary and specific to each equipment manufacturer, leading to vendor lock-in scenarios in which network operators are dependent on a single supplier for equipment and updates, thereby stifling competition and innovation within the industry.

\subsubsection{Centralized and Cloud RAN}
The idea of centralized/cloud RAN (C-RAN), as shown in Fig. \ref{fig:cran}, is to group the BBUs of multiple BSs in a single location, referred to as a BBUs hotel or pool of BBUs. When the BBUs are located at a physical site, the network is called a centralized RAN, whereas in the cloud, it is called a cloud RAN. C-RAN emerged as a solution to the cost, space, and maintenance challenges in D-RAN by leveraging cloud technologies to centralize baseband processing functions and achieve resource efficiency and scalability. This approach enables dynamic resource allocation based on demand, optimizing network utilization, accommodating varying workloads, and facilitating the seamless introduction of new services and capabilities without requiring extensive hardware modifications or upgrades.

\begin{figure}[!htbp]
\centering
\includegraphics[width=0.9\linewidth]{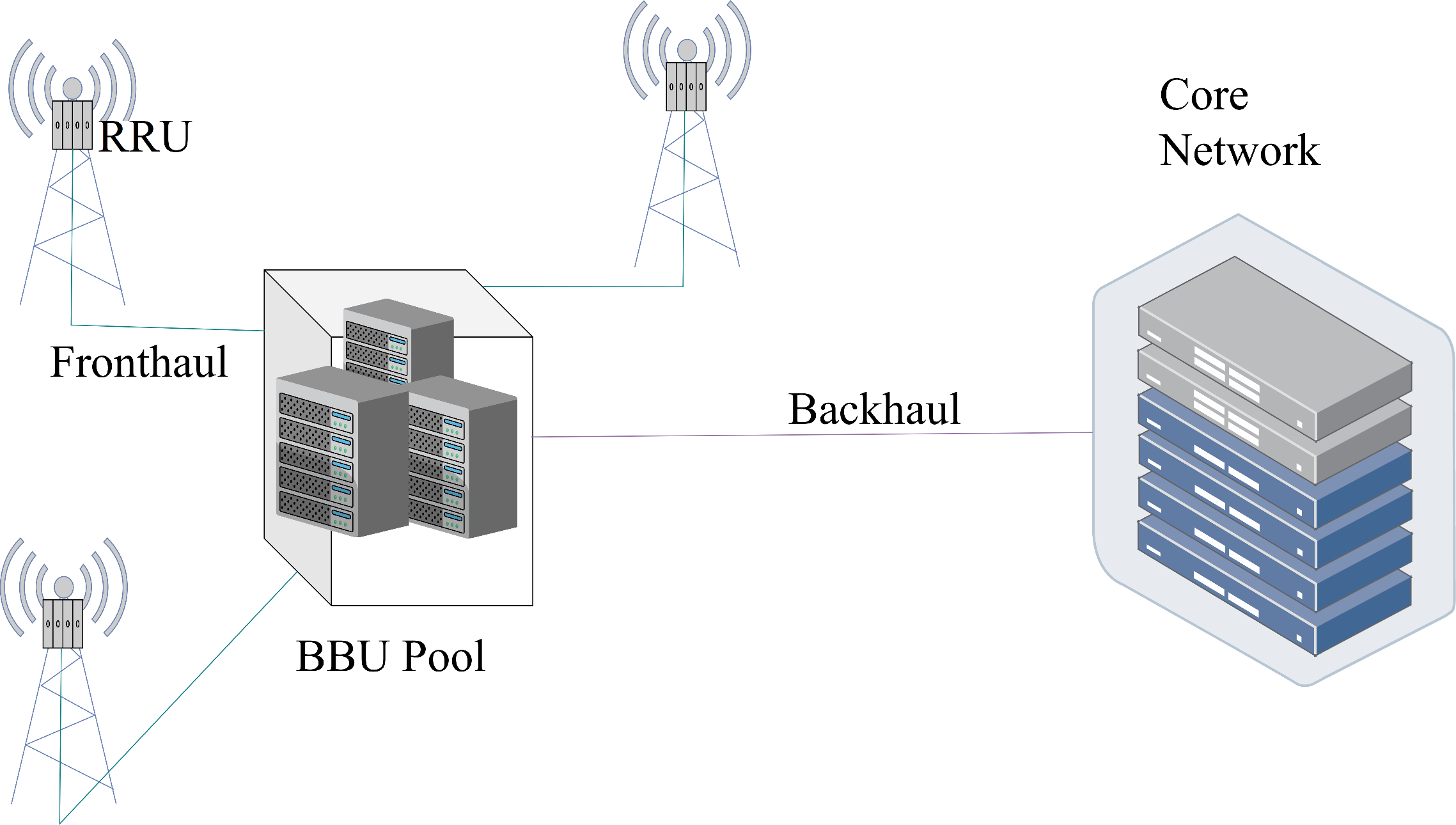}
\caption{Architecture of Centralized RAN}
\label{fig:cran}
\end{figure}

The interface between the BBU pool and the RRUs is achieved through the fronthaul connection, responsible for conveying user data, control signals, and baseband information between the BBU and RRUs with strict requirements for bandwidth and latency. The adoption of fiber cable with high bandwidth and low-latency over the CPRI protocol as transport technologies is essential in meeting these demands, ensuring that the separation of baseband processing from the radio elements at distances up to 15-20Km does not compromise the network's performance or user experience \cite{Agrawal_2017, 6897914}.

Despite the great advantages, C-RANs depend on a high-performance fronthaul interface, which introduces new complexities in network design and operation, requiring sophisticated synchronization mechanisms and advanced error correction techniques to mitigate latency and ensure data integrity across the network. Moreover, the centralization of baseband processing raises concerns regarding fault tolerance and resilience, as the consolidation of resources could potentially create single points of failure that might impact network reliability \cite{7986431}.

\subsubsection{Virtual RAN}

 Virtual RAN (vRAN) extends the principles of centralization and resource pooling inherent in C-RAN by leveraging virtualization technologies to abstract the baseband processing functions from the underlying hardware. The virtualization is the abstraction of the hardware resources to decouple NFs from proprietary hardware, allowing these functions to run as software instances on commodity servers, as illustrated in Fig. \ref{fig:vran}. This fundamental shift from a hardware-centric to a software-defined network (SDN) brought higher levels of flexibility, scalability, and efficiency beyond C-RAN architectures, paving a new trajectory to next-generation mobile networks \cite{8723481}.

 The benefits of vRAN include a significant degree of operational flexibility, allowing network operators to swiftly respond to changes in network conditions and demand patterns thanks to the ability to instantiate, scale, or decommission NFs virtually, without the need for physical interventions \cite{10464310}. Agility is particularly crucial for 5G applications, where network slicing and other advanced functionalities demand a highly adaptable network infrastructure. Moreover, the SDN approach reduces reliance on specialized hardware, leading to substantial CapEx and OpEx savings, and resource virtualization inherently promotes dynamic optimization of resources like CPU, memory, and storage based on demand, preventing over-provisioning and under-utilization \cite{9684900}.

The stringent performance requirements inherent in recent network generations, particularly in terms of latency and throughput defies the capabilities of vRAN. The virtualization layer introduces additional complexity and potential processing overhead, which is a concern for services with low-latency and high-reliability requirements, such as ultra-reliable low-latency communication (URLLC) and enhanced mobile broadband (eMBB) in 5G \cite{7324637}. Secondly, the dynamic nature of virtualized NFs requires sophisticated orchestration capabilities to ensure seamless operation, optimal resource allocation, and fault tolerance, resulting in compounded complexity and challenges in terms of standardization and compatibility. From a security perspective, disaggregation of VNFs and the reliance on shared infrastructure and cloud platforms introduce new vulnerabilities and attack vectors, necessitating robust security mechanisms and protocols to protect network integrity and user data.

Just like the previous shifts in RAN architectures throughout different network generations, the advent of O-RAN is triggered by the need of better flexibility and adaptability, cost effectiveness, higher security levels, and network automation and intelligence \cite{7355588,6568922}.

\begin{figure}[!htbp]
\centering
\includegraphics[width=0.9\linewidth]{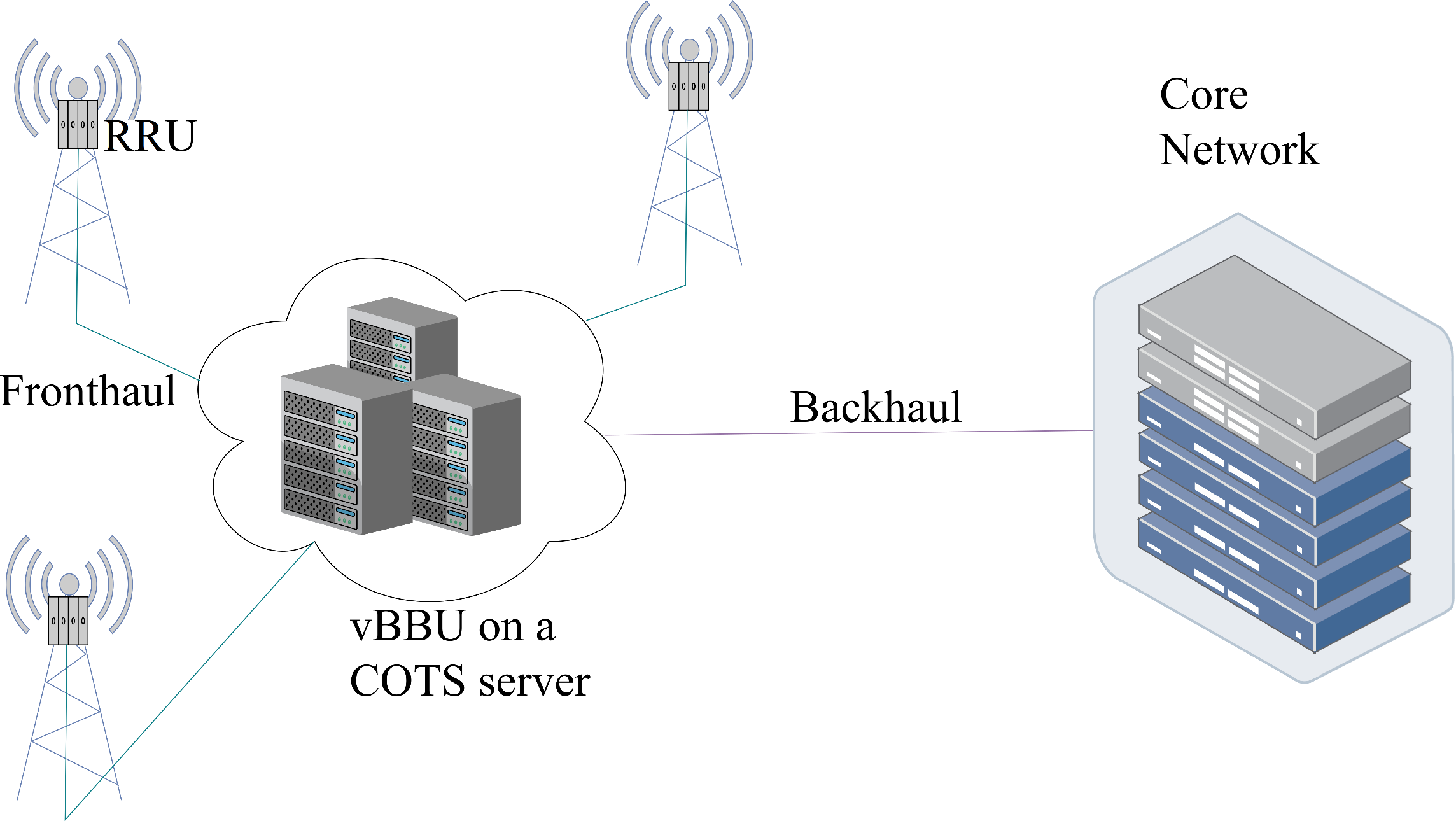}
\caption{Architecture of virtual RAN}
\label{fig:vran}
\end{figure}

\subsection{Foundational principles of O-RAN}
As 5G continues to demonstrate its effectiveness, traditional network architectures are failing to support stricter services requirements,  forcing vendors and mobile network operators to consider O-RAN as a new architectural paradigm \cite{Wypior_Klinkowski_Michalski_2022}. The expectations for 5G and beyond networks are substantial, including ultra-high reliability and low latency for mission-critical services, better connectivity for massive machine type communications (mMTC) to meet IoT applications needs, and high throughput for eMBB applications such as video surveillance, teleconferencing, and remote surgery should also be supported \cite{10073507}. To effectively address these strict requirements and high expectations the O-RAN concept was built around  four foundational principles\cite{aly_s_abdalla_toward_2022}:

\subsubsection{Disaggregation}
Disaggregation in O-RAN refers to the segmentation of the RAN into standardized and interoperable components. It splits the traditional tightly integrated and single-vendor RAN setup into different units, allowing hardware and software from different vendors to work together, promoting competition, innovation, and reducing costs \cite{10024837}. As illustrated in Fig. \ref{fig:O-RAN-disegragation}, the RAN architecture becomes modular by being broken down into distinct, independently managed functional elements, namely the O-RAN compatible RU (O-RU), the O-RAN compatible distributed unit (O-DU), and the O-RAN compatible centralized unit (O-CU). Each unit hosts specific NFs and can be independently sourced, upgraded, or replaced. Units also seamlessly integrate with one another through the principles of openness and interoperability \cite{10024837}.
\begin{figure}[!htbp]
    \centering
    \includegraphics[width=0.9\linewidth]{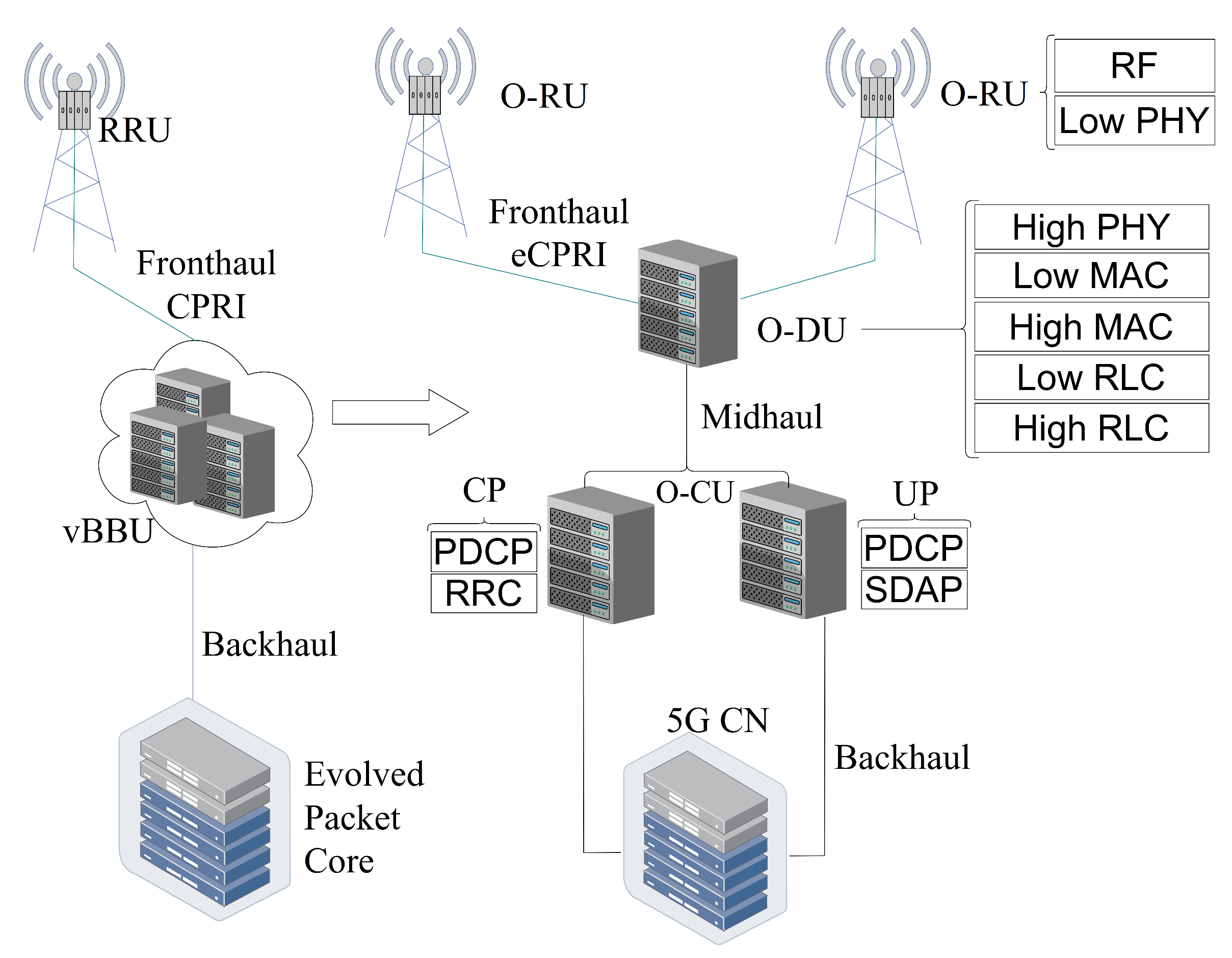}
    \caption{Network architecture disaggregation in O-RAN}
    \label{fig:O-RAN-disegragation}
\end{figure}

\subsubsection{Virtualization} Although the concept of virtualization is not new in RAN architectures, it serves as a primary key principle in O-RAN architecture, providing great flexibility in RAN management by allowing migration of the NFs implementation from proprietary and vendor-specific hardware to COTS platforms using virtual machines or containerized applications. The abstraction of COTS hardware resources through virtualization makes it possible to deploy O-DUs and O-CUs as virtual machines or containers. It simplifies the scaling up and down based on the network traffic demand, helps to fully automate the network slicing services such as instantiation, scaling, and continuous integration and deployment \cite{10248352}. 

\subsubsection{Openness and interoperability}
The openness principle advocates for open interfaces and protocols between different modular elements of the RAN to ensure they can efficiently interoperate within the network infrastructure regardless of the manufacturer. By promoting open interfaces, O-RAN ensures that operators can mix and match hardware and software from different suppliers (for example O-RU from one vendor, the O-DU from another vendor, and the O-CU from a third vendor), fostering a competitive and diverse ecosystem. The openness principle aims to drive innovation and prevent vendor lock-in, which offers operators more flexibility in building and managing their networks \cite{Thiruvasagam_2023}.

\subsubsection{Intelligence and programmability}
 The possibility of managing and optimizing radio resources through third-party applications is another peculiarity in O-RAN architecture. These applications, called xApps (for real time operations) and rApps(for non real-time operations) are deployed in the RICs to close-loop control the RAN functions through open APIs. The RICs facilitate control mechanisms that continuously monitor, analyze, and optimize RAN parameters and network functions in both near-real-time (1ms to 10ms) and non-real-time (up to 1s), thereby enhancing network performance and adaptability. The programmability refers to the ability to configure and adapt policies using AI/ML techniques \cite{giannopoulos_supporting_2022, 10506767}.

\subsection{Comprehensive analysis of O-RAN architecture}
The development of O-RAN architecture began in 2018 when a consortium of vendors and operators, known as O-RAN Alliance, was established to develop and adopt standard specifications for making next-generation wireless access networks disaggregated, virtualized, open, intelligent, and interoperable \cite{10024837}.
The foundational architecture is formally defined in the O-RAN Alliance's technical specification, which outlines the principles and components of an open, intelligent, and multi-vendor RAN \cite{oran2018whitepaper}.
It complies with the 3GPP 5G RAN architecture, which splits the base-band processing stack functions across three logical nodes, the CU, DU, and RU to meet specific operational requirements. The different options for splitting the stack functions are called functional splits and were initially outlined in 3GPP Release 14 and further defined in 3GPP Release 15 \cite{3gppTR21914Rel14, 3gppTR21915Rel15}. They range from low-level splits to high-level splits that allow for more complex processing at the edge of the network. O-RAN Alliance adopted the split 7.2x, illustrated in Fig. \ref{fig:O-RAN-disegragation}, as the standard for the O-RAN architecture to allow a more dynamic allocation of resources and functions across the network.

This section delves into the primary components of O-RAN logical architecture as illustrated in Fig. \ref{fig:open-ranFunc}, including the O-RU, O-DU, O-CU, RICs, open interfaces, and service management and orchestration (SMO).
\begin{figure}[!htbp]
    \centering
    \includegraphics[width=0.9\linewidth]{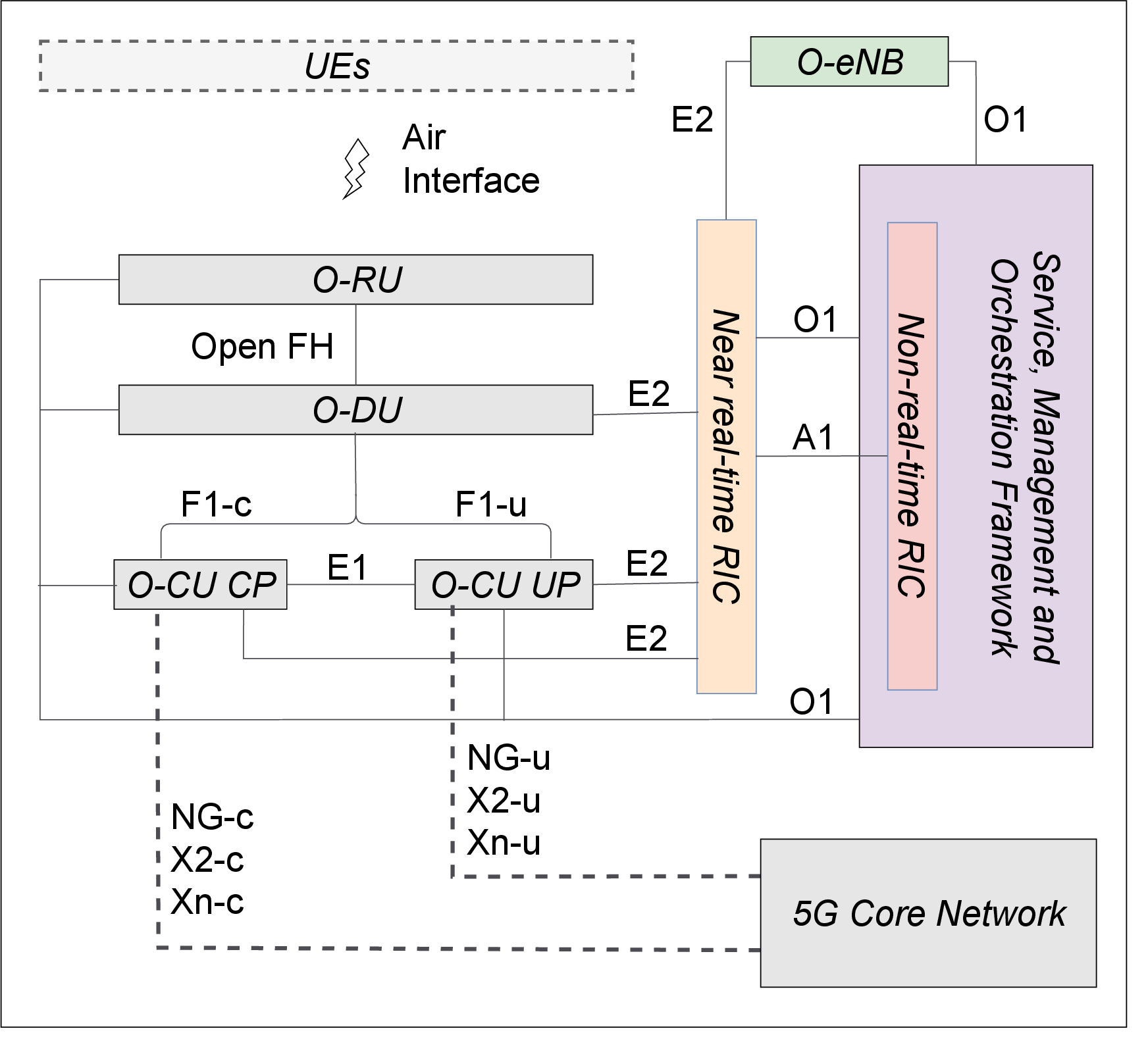}
    \caption{Functional architecture of O-RAN, illustrating the disaggregation of baseband processing elements into distinct units.}
    \label{fig:open-ranFunc}
\end{figure}
}

\subsubsection{Open Radio Unit (O-RU)}
The O-RU handles RF processing and transmission, directly interfacing with antennas. Its design incorporates a modular RF front-end for signal amplification and filtering, along with base-band processing units for tasks like modulation/demodulation. A standardized and open interface, known as the fronthaul, connects the O-RU to the O-DU over the enhanced CPRI (eCPRI) protocol \cite{8736787}, ensuring vendor-agnostic interoperability and flexibility in network deployment. O-RAN handles the radio frequency and the lower PHY layer functions, and plays a vital role in enhancing network coverage and capacity while supporting various frequency bands and technologies.

\subsubsection{Open Distributed Unit (O-DU)}
The O-DU is a logical node responsible for real-time ($<$10ms), Layer 2 baseband processing, and its hosted functions primarily include the physical layer high-PHY, medium access control layer, and the radio link control layer, which are critical for time-sensitive operations \cite{oran2025usecasesdetail, oran2025usecasesanalysis}. Its key responsibilities are as follows:

\noindent \textit{Real-Time Processing Unit:} executes radio resource scheduling, radio link control, and hybrid automatic repeat request (HARQ) functions.

\noindent \textit{Resource Management:} allocates radio resources dynamically based on demand.

\noindent \textit{Interface to O-RU and O-CU:} connects to the O-RU via the standardized Open fronthaul interface and to the O-CU via the F1 interface. This facilitates the low-latency data flow and control signaling required for coordinated network operation.

\subsubsection{Open Centralized Unit (O-CU)}
O-CU handles non-real-time processing tasks and is responsible for higher-layer functions. {\color{black}It is further disaggregated into a Control Plane (O-CU-CP) and a User Plane (O-CU-UP) \cite{oran2025usecasesdetail, oran2025usecasesanalysis}, to enable independent scaling and evolution of each plane:

\noindent \textit{O-CU-Control Plane (O-CU-CP):} hosts the radio resource control (RRC) and packet data convergence protocol (PDCP) control part. It is responsible for signaling, mobility management, session management, and controlling the O-CU-UP.

\noindent \textit{O-CU-User Plane (O-CU-UP):} hosts the User Plane part of the PDCP protocol and the service data adaptation protocol (SDAP). Its primary function is to handle and route user data traffic, ensuring efficient data flow with QoS enforcement.

The O-CU-CP and O-CU-UP communicate with each other via the E1 interface. The O-CU connects to the O-DU using the F1 interface (F1-C for control and F1-U for user data), facilitating the critical data flow and control signaling between these distributed units.}
\\
By centralizing these functions, the O-CU can optimize resource utilization and improve overall network efficiency. The separation of the O-CU from the O-DU allows for a more flexible architecture, enabling operators to deploy resources based on specific service requirements and traffic patterns.

\subsubsection{RAN intelligent controllers (RICs)}
RICs are pivotal in the O-RAN architecture, and they provide advanced data-driven analytics and ML capabilities. {\color{black}The architecture features two types of RICs: the non-real-time (Non-RT) RIC and the near-real-time (Near-RT) RIC hosted in the SMO and operating on timescales greater than 1 second, manages high-level policies and ML model training, and communicates with the Near-RT RIC via the A1 interface \cite{Hassouna_2025, CSRIC_O_RAN_2022, ORAN_NGRG_2025, FCC_TAC_6G_2025}.} The Near-RT  RIC, positioned closer to the network edge and operating between 10 ms and 1 second, controls RAN elements through the E2 interface to apply policies and perform near-real-time optimization.
RICs enable operators to implement intelligent resource management strategies, enhance user experience, and adapt to changing network conditions. By leveraging AI and ML, RICs can predict traffic patterns, optimize resource allocation, and improve overall network reliability.

\subsubsection{Open interfaces}
Open interfaces are a fundamental element of O-RAN architecture, promoting interoperability and modularity among different network components. {\color{black}The O-RAN Alliance specification defines a comprehensive suite of standardized interfaces, such as A1 (between Non-RT RIC and Near-RT RIC), E2 (between Near-RT RIC and its controlled functions), O1 (for management), and the Open Fronthaul (between O-DU and O-RU) \cite{oran2018whitepaper, 10024837}.} Open interfaces serve as communication methods between the RAN components, allowing for seamless integration and interaction among various vendors' equipment, by ensuring that components can work together regardless of the manufacturer. By standardizing interfaces, O-RAN reduces the complexity of network integration, fosters innovation, and enables operators to mix and match components from various suppliers, enhancing flexibility and reducing costs.

\subsubsection{Service Management and Orchestration (SMO)}
{\color{black} 
The SMO framework in the O-RAN architecture is a modular, cloud-native framework designed to manage, automate, and orchestrate the lifecycle of disaggregated RAN functions across multi-vendor components. According to \cite{ORAN_SMO_Architecture_2024, oran2018whitepaper, 10024837}, the SMO provides a broad set of functions that span operations, administration, and maintenance (OAM), including performance assurance, fault supervision, and provisioning. It also supports rApp lifecycle management within the Non-RT RIC, exposes topology and inventory information, and offers service and data management interfaces. Furthermore, the SMO is responsible for orchestrating O-Cloud resources, enabling network slicing, performing traffic analytics, and supporting service assurance. In addition, it governs data and service exposure through standardized interfaces to ensure interoperability across multi-vendor and heterogeneous O-RAN deployments. Moreover, the SMO supports AI/ML workflows, allowing for the onboarding, training, and deployment of models that optimize spectrum usage, resource allocation, and models for security. These models are validated through rigorous testing pipelines before being deployed into RICs.} Its key functionalities are:
\\
\textbf{Management Functions:} Oversee the deployment, configuration, and optimization of network resources.
\\
\textbf{Orchestration Layer:} Coordinates the interactions between different RAN components.
\\
\textbf{Monitoring and Analytics:} Provides insights into network performance and service quality.
\\
By integrating SMO into the O-RAN architecture, operators can achieve greater agility, reduce operational costs, and enhance service delivery.
\\

{\color{black}
\subsubsection{Data management in O-RAN}
O-RAN is designed to be mainly data-driven, with closed-loop controls depending on an effective and continuous lifecycle of data collection, management, and utilization through open interfaces and RICs. Data is gathered from various components of the RAN and the O-RAN infrastructure itself through standardized open interfaces.
\begin{itemize}
    \item \textit{E2 Interface:} it serves as the primary conduit for near-RT telemetry from RAN components. Data includes user-level and cell-level key performance measurements (KPMs), event triggers like handover requests, and configuration states.
    \item \textit{O1 Interface:} used for non-RT management data, including performance assurance reports, fault supervision alerts, configuration data, and trace information from all O-RAN managed elements.
    \item \textit{A1 Interface:} carries enrichment information and policies from the non-RT RIC to the near-RT RIC, which can include external data or aggregated analytics not directly available from the RAN, hence enhancing decision-making by providing broader insights beyond raw telemetry.
\end{itemize}

Collected data is streamed to the RICs and aggregated into centralized repositories or data lakes within the SMO framework for large-scale analysis. The raw data then undergoes preprocessing, including formatting, normalization, scaling, and dimensionality reduction with autoencoders, to make it suitable for analysis and model training. The processed data is then maintained in structured storage systems. In the near-RT RIC, the shared data layer (SDL) and network information base (NIB) provide a low-latency, shared database for xApps to store and access RAN context, including instance-connected users and node lists. Meanwhile, in the SMO/non-RT RIC, data lakes store vast historical datasets for offline training, validation, and long-term trend analysis.}

{\color{black}
\section{ ML in O-RAN}
{\color{black}
\subsection{ML: A Brief Overview} 

% One field of AI is ML, which focuses on developing algorithms and models that enable computers to acquire knowledge and discover patterns in data, allowing them to predict or make decisions based on that information\cite{wang_machine_2021,utmal_machine_2021}. Furthermore, it has been shown in \cite{sadiq_review_2022} that ML is a methodology that facilitates knowledge acquisition and analysis of historical data.

The versatility of ML makes it increasingly relevant in computer networks, particularly in the context of emerging O-RAN architectures. While ML has demonstrated significant potential in fields such as healthcare—for disease detection, diagnosis, and prognosis \cite{javeed_machine_2023}- and finance—for fraud detection, risk assessment, and forecasting \cite{mahdi_software_2021}- its application in networking enables automation, optimization, and enhanced decision-making. In particular, ML improves network security by supporting intrusion detection, malware analysis, and cyber threat identification \cite{ali_effective_2023,vaid_intrusion_2021,ferrag_edge-iiotset_2022}. In computer networks, ML has become essential for achieving objectives such as quality of experience (QoE) management, traffic classification, and resource optimization. For instance, in SDN-based networks, the separation of control and data planes provides flexibility that ML can exploit to dynamically analyze traffic patterns, user behavior, and network conditions to optimize routing, improve QoE, and manage resources efficiently \cite{wang_machine_2021}, \cite{samadi_machine_2022}. Moreover, ML algorithms enable automated network traffic classification, grouping flows based on protocols, applications, services, or content, which facilitates monitoring, security enforcement, and QoS management \cite{vulpe_machine_2021}. In wireless networks, ML has been successfully applied to predict network usage patterns and optimize bandwidth reservation by analyzing historical traffic data \cite{genda_-demand_2021}. These capabilities directly support the key objectives of O-RAN, including intelligent spectrum management, adaptive resource allocation, and enhanced security, highlighting the crucial role of ML in enabling more efficient, secure, and autonomous next-generation networks.}

\subsection{AI in O-RAN: Types and Transformative Impacts of ML}

ML plays a vital role in O-RAN due to its ability to enhance network performance, automate processes, and address complex challenges, such as accelerating resource arbitration procedures that govern how resources are allocated in 5G RAN slicing, ultimately improving allocation efficiency \cite{yarkina_performance_2024}. Moreover, ML has become crucial in the intelligent resource management of RAN slicing, which enhances the overall network performance and facilitates the improvement of network resources \cite{azimi_applications_2022}. Hence,  the application of ML in O-RAN opens up opportunities for building networks that can adaptively manage and optimize their own performance \cite{iturria-rivera_multi-agent_2022}.

In addition, ML and AI are essential components in the implementation of O-RAN as they offer intelligent and adaptable features to the network structure \cite{giannopoulos_across_2023}. An intelligent O-RAN framework that uses game theory and ML demonstrates the importance of ML in reducing complexity and facilitating intelligent network operations \cite{abedin_elastic_2022}. Moreover, the increased automation and efficiency of O-RAN services are also inseparable from the role of the ML approach, such as predicting the amount of resources required by each network slice to meet the Service Level Agreement (SLA). This enables automated, proactive, and adaptive network management, as the system will automatically and proactively predict changes in resource requirements and adapt to changes in network conditions and user needs \cite{thaliath_predictive_2022}.

\iffalse
Moreover, the advancement of predictive closed-loop service automation in O-RAN-based network slicing emphasizes using ML approaches to manage the diversity of 5G wireless networks effectively, including network slicing. It enables proactive and adaptable network management \cite{thaliath_predictive_2022}.
\fi

\subsubsection{SL in O-RAN} 

SL is a form of ML in which algorithms are trained with labeled data to generate either predictions or make decisions, and classification \cite{cryst14100830,9350377}. The training process involves utilizing input-output pairs, where the learning algorithm establishes a mapping between the inputs and outputs. The objective for the algorithm is to possess the capability to provide forecasts or determinations when novel or previously non-existent data is introduced to it \cite{tyagi_chapter_2022}. 

SL in O-RAN  significantly impacts network management and optimization aspects. By leveraging SL algorithms, O-RAN architectures can enhance resource allocation, anomaly detection, intelligence support, mobility management, network slicing, and rogue BS detection. For instance, \cite{makhlouf_optimized_2023} has shown that DL models can be effectively utilized to allocate resources efficiently within the O-RAN architecture. This optimization leads to improved network performance, reduced latency, enhanced user experiences, and better utilization of available resources. 

Furthermore, \cite{haryo_prananto_o-ran_2022} provides insights into the benefits of SL in O-RAN for mobile mobility management, potentially highlighting how SL contributes to optimizing the handover process and improving overall mobility management within the network.
Additionally, in terms of network security and reliability, SL methods were leveraged to identify and classify near real-time interference in 5G New Radio (NR) with the help of Bayesian inference to enhance the security elements of O-RAN \cite{jere_bayesian_2023}. SL can also be utilized to detect anomalies in 5G O-RAN architecture by identifying and addressing irregularities in the network \cite{alves_machine_2023}, thus contributing to the overall security and stability of the O-RAN environment. Furthermore, \cite{huang_developing_2023} identifies unauthorized BS in O-RANs supporting Software-Defined Radio (SDR) by using xApps generation that applies ML methods to improve network security and reliability. 
It highlights the critical role of SL in improving network security and stability by enabling proactive anomaly detection and mitigation. This occurs through efficiently identifying and addressing possible security risks to ensure network integrity.

Moreover, SL plays a crucial role in network slicing within O-RAN designs. By utilizing SL algorithms, O-RAN systems can optimize network slicing processes for specific applications, such as smart grid applications \cite{Ho_Collaborative_2023}. This optimization ensures that network resources are efficiently allocated to meet the diverse requirements of different services, enhancing the overall flexibility of O-RAN deployments. Furthermore, SL contributes to intelligence support in disaggregated O-RAN networks by implementing SL-based algorithms for tasks such as cell traffic prediction \cite{giannopoulos_supporting_2022} and enhancing traffic prediction capabilities \cite{Moreira_Baseband_2022}.
In summary, SL is a cornerstone of O-RAN, supporting cell traffic prediction, anomaly detection, intelligent decision-making, cellular mobility management, and network slicing.

\subsubsection{UL in O-RAN}
UL is a fundamental concept in ML that operates on unlabeled input data \cite{zhang_research_2022}. Thus, the model can explore and gain insights from the data independently, allowing the model to identify patterns, structures, and relationships in input data without external guidance \cite{chen_review_2021,8537776}. UL is also essential in enabling network automation by analyzing and understanding the behavior of different network slices and resource requirements. Using its algorithms, such as clustering, network operators can gain insight into traffic patterns, resource utilization, and performance metrics across different network slices without requiring labeled training data. It enables automatic identification of similarities, anomalies, and optimal resource allocation strategies based on the intrinsic characteristics of the data.

UL is quite beneficial in O-RAN for the adaptive retraining of AI/ML models for Beyond 5G networks,  a predictive approach, which leverages UL techniques to enhance the QoS in computer networks is involved \cite{gudepu_adaptive_2023}. This approach aims to continuously improve the performance of AI/ML models by predicting and adapting to network dynamics without the need for labeled training data, enabling    AI/ML models to autonomously identify patterns and trends in network behavior.

Furthermore, in terms of AoP (Age of Processing) towards offloading autonomous vehicle data to the edge cloud in Multiple Radio Access Technologies (MultiRAT) O-RAN, UL algorithms are employed to facilitate efficient data processing tasks  \cite{ndikumana_age_2022}. The primary goal is to ensure seamless and reliable communication for autonomous vehicles by dynamically managing the processing and routing of data. UL algorithms enable the system to analyze and categorize data traffic patterns, identify processing requirements, and make real-time decisions regarding data offloading without the need for explicit supervision. This approach ultimately aims to optimize the utilization of network resources, minimize latency, and enhance the overall communication experience for autonomous vehicles operating within the O-RAN framework.

\subsubsection{RL in O-RAN}
RL is a type of adaptive ML where agents learn to make decisions that maximize long-term rewards based on interaction with the environment \cite{moudoud_empowering_2023}. In the O-RAN context, RL has been used for various purposes, such as resource allocation, distributed intelligence, and hosting AI/ML workflows. Most previous studies use RL to optimize resource allocation, which is still developing in O-RAN research. Previous studies were conducted based on different needs for different services, such as high peak rates for eMBB, and low delays in URLLC and 5G devices that require mass connections for mMTC \cite{Cheng_Reinforcement_2023}. The RL approach used for resource blocks (RBs) selection for each network traffic based on throughput as a key performance indicator (KPI), uses SARSA on-policy differential semi-gradient \cite{Mungari_RL_Approach_2021}, while \cite{Kouchaki_Actor_2022} utilized Advantage Actor-Critic (A2C) and Proximal Policy Optimization (PPO) to allocate Resource Block Group (RBG). Furthermore, some studies not only use RL, but also combine it with Transfer Learning (TL) \cite{Nagib_Safe_2024} and FL \cite{firouzi_5g-enabled_2022}. The work in \cite{Nagib_Safe_2024} leveraged the Hybrid Policy Transfer approach, which consists of Policy Reuse and Policy Distillation, to allocate physical resource blocks (PRB) in certain slices to meet SLA requirements. The use of RL in O-RAN, apart from being more adaptive, can also be safer and able to reduce costs, as the time required for the slicing process can be reduced.

\subsubsection{FL in O-RAN}
ML also has a paradigm called FL that uses decentralized techniques that are different from conventional techniques. This method no longer requires data to be centralized in one location, as training is performed in different locations. This has significant advantages in terms of security and accessibility \cite{10622458,10592377, abouaomar_federated_2023}. Hence, combining FL with O-RAN can be useful in handling sensitive user data and maintaining data security \cite{Noureen_Federated_2023}. Furthermore, \cite{firouzi_5g-enabled_2022} proposed a three-layer, client-edge-cloud, FL-based architecture that is optimized using RL in client selection and resource allocation. The approach is capable of handling the challenge of ensuring optimal device selection and resource allocation decisions online. \cite{firouzi_5g-enabled_2022} proposes Federated Learning implemented on a three-layer "client-edge-cloud" architecture to update local parameters "client-edge" and global aggregation "edge-cloud" and leverage reinforcement learning for user selection on each FL task. They state that it can balance performance and learning costs with the proposed framework.

{\color{black}
As shown in Table \ref{tab:ml_in_O-RAN}, ML is being increasingly integrated into O-RAN through a variety of learning approaches, highlighting the substantial potential of ML-enabled O-RAN architectures. This table systematically summarizes the types of ML employed, the specific algorithms applied, and the tasks targeted within O-RAN. Notably, RL emerges as the most widely adopted technique, particularly for resource management, due to its natural alignment with the adaptive decision-making requirements and highly dynamic environment of O-RAN. In contrast, SL and UL are less frequently applied for resource management, control, and optimization, largely because they rely on labeled or static datasets, which are challenging to obtain in real-world O-RAN deployments.

As illustrated in Fig. 8, research on ML in O-RAN has been steadily increasing. In 2021, only 7.42\% of studies focused on ML, but this interest grew rapidly in subsequent years: 20.14\% in 2022, 22.26\% in 2023, and 24.38\% in 2024. Preliminary data for 2025 indicates a further rise to 25.80\%, demonstrating a continuing upward trend. These figures underscore the growing importance of intelligent, data-driven methods in O-RAN development.
Furthermore, Fig. \ref{fig:pict2} provides a breakdown of the types of ML applied within O-RAN. RL stands out as the dominant approach, featuring in approximately 63\% of studies. This emphasis reflects RL’s ability to continuously learn and adapt through interaction with the network, making it particularly well-suited for complex tasks such as resource allocation, RAN slicing, scheduling, and session management, where real-time adaptability is critical for optimal performance.

}

\begin{figure}[h!]
\centering
\includegraphics[width=3.5in]{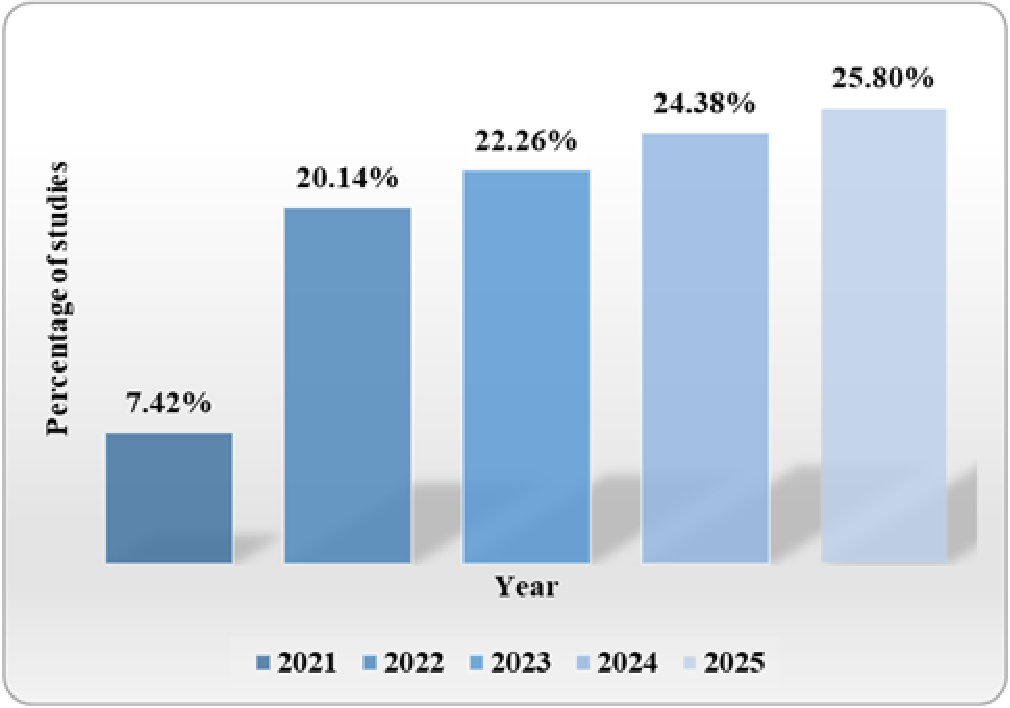}
\caption{Percentage of ML research in O-RAN}
\label{fig:pict1}
\end{figure}

\begin{figure}[h!]
\centering
\includegraphics[width=3.0in]{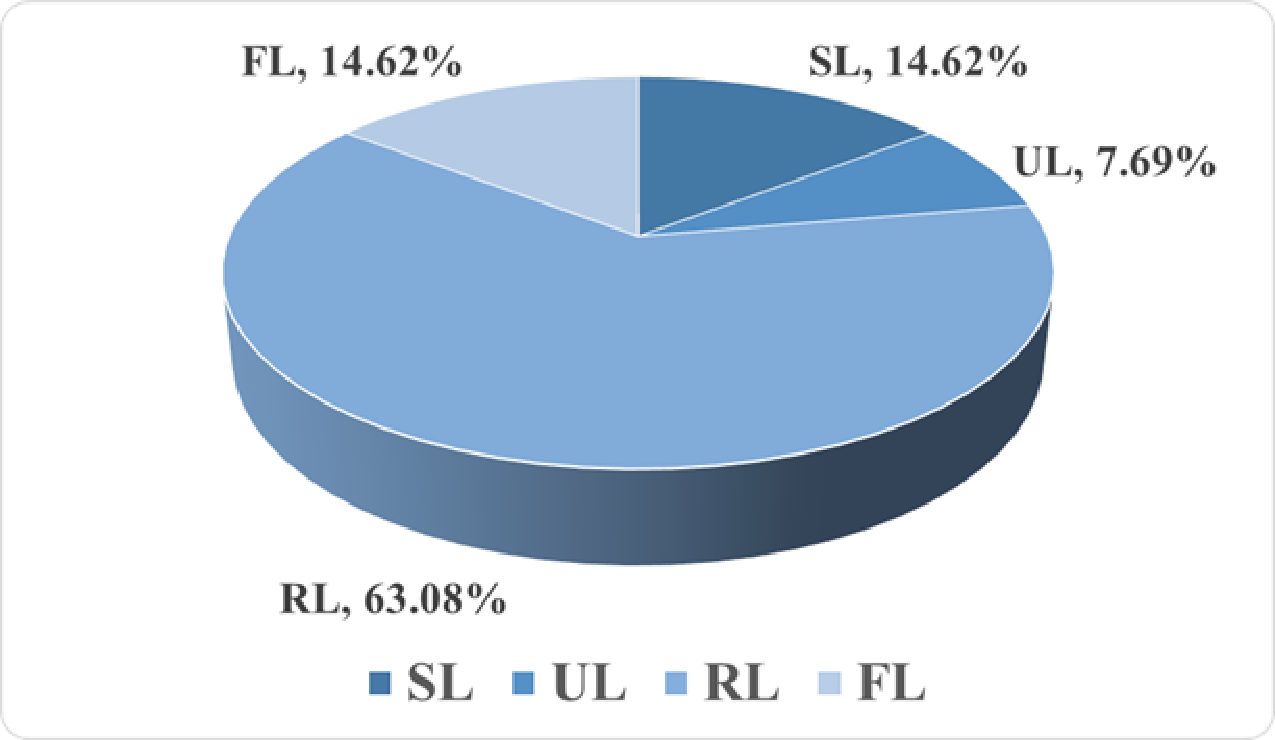}
\caption{Percentage of ML research pertaining to O-RAN categorized by type}
\label{fig:pict2}
\end{figure}

\begin{table*}[htbp]
\centering
\caption{ML Utilization in O-RAN}
\label{tab:ml_in_O-RAN}
\begin{tabular}{|p{0.65cm}|p{3.5cm}|p{7.5cm}|p{3.4cm}|}
\hline
\multicolumn{1}{|c}{\textbf{Type}} & \multicolumn{1}{|c|}{\textbf{Algorithm}} & \multicolumn{1}{c|}{\textbf{Task}} & \multicolumn{1}{c|}{\textbf{Ref}}\\
%Type & Algorithm & Task \\ 
\hline
\multicolumn{1}{|c|}{\textbf{\multirow{10}{*}{SL}}} & FFNN,RNN & Traffic prediction and VNF baseband allocation & \cite{Moreira_Baseband_2022} \\
\cline{2-4}
 & ARIMA & Predicting the scaling of the number of VNFs & \cite{Ali_Proactive_2024} \\
\cline{2-4} 
 & \multirow{4}{*}{LSTM} & RAN Slicing &  \cite{Ho_Collaborative_2023}\\
\cline{3-4} 
 & & Predicting the scaling of VNFs' number; Handover organizing & \cite{Ali_Proactive_2024}\\
\cline{3-4}
 & & Anomaly Detection & \cite{rumesh_federated_2024} \cite{issa_evaluation_2025} \\ 
\cline{3-4}
 & & Network Energy Saving & \cite{kim_open_2025} \\ 

% \cline{2-4} 
\cline{2-4} 
 & GNB,NN,KNN & Predicting communication compatibility times; Attack detection & \cite{Rastogi_Intelligent_2023} \cite{huang_developing_2023}\\
\cline{2-4}
 & RF, SVM & Attack detection; User classification & \cite{huang_developing_2023}\cite{Anand_Machine_2023}\\
\cline{2-4}
 & GBT,CNN & User classification & \cite{Anand_Machine_2023} \\ 
\cline{2-4}
 & LSTM, XGBoost & Cell throughput prediction & \cite{nassar_cell_2024} \\ 
\cline{2-4}
 & HGBoost, KNN, NN, SVM & Predicting Metrics for Control and Management & \cite{herrera_enabling_2024} \\ 
\cline{2-4}
 & SCL & Network slice prediction & \cite{he_contrastive_2025} \\ 
 \cline{2-4}
 & CNN (Incremental Learning) & Anomaly Detection & \cite{gain_fortifying_2025} \\ 
  \hline
 
\multicolumn{1}{|c|}{\textbf{\multirow{5}{*}{UL}}} & LSTM & Handover organizing \& MCS selection & \cite{9627159}\\ 
\cline{2-4} 
 & IF, LSTM-AutoEncoder & Anomaly Detection & \cite{Başaran_Deep_2023} \cite{rumesh_federated_2024}\\ 
\cline{2-4} 
 & K-Means & Traffic steering and load balancing & \cite{Ntassah_xApp_2023}\\ 
\cline{2-4} 
 & DBSCAN & Resource allocation & \cite{Rezazadeh_Specialization_2023} \\ 
 \cline{2-4} 
 & UCL &  Network slice prediction & \cite{he_contrastive_2025} \\ 
\hline

\multicolumn{1}{|c|}{\textbf{\multirow{46}{*}{RL}}} & \multirow{3}{*}{PPO} & VNF scaling and placement & \cite{Ali_Proactive_2024}\\
\cline{3-4}
 & & Radio Resource Allocation & \cite{Pandey_DeepReinforcement_2023}\cite{Hammami_On_Policy_2022}\\
\cline{3-4}
 & & Controlling cell activation and deactivation & \cite{bordin_demo_2025} \cite{bordin_design_2025}\\
 
\cline{2-4}
 & \multirow{4}{*}{Deep Q-Learning} & Offloading and fronthaul routing & \cite{Ndikumana_Federated_2023}\\
\cline{3-4}
 & & Power adjusment & \cite{Vankayala_Reinforcement_2022}\\ 
\cline{3-4}
 & & VNF allocation & \cite{Tamim_ALAP_2023}\\
\cline{3-4}
 & & Energy Efficiency Optimization & \cite{hoffmann_energy_2024}\\
 
\cline{2-4}
 & \multirow{2}{*}{DDQN} & Radio Resource Allocation &  \cite{Wang_Resource_2023}\cite{firouzi_5g-enabled_2022}\cite{Rezazadeh_Multi_2023}\cite{Filali_Communication_2023}\cite{10433640}\\
\cline{3-4}
 & & VNF scaling and placement & \cite{Rezazadeh_Specialization_2023}\\ 
 
\cline{2-4}
 & \multirow{2}{*}{DDPG} & Cache Repository Selection & \cite{Tran_Joint_2022}\\ 
\cline{3-4} 
 & & Radio Resource Allocation & \cite{Lotfi_Evolutionary_2022} \cite{wang_energy-efficient_2024} \cite{RAGO_A_Tenant_2022}\\
 
\cline{2-4}
 & \multirow{3}{*}{Q-Learning} & Handover Management & \cite{Ortiz_Q-learning_2023}\\ 
\cline{3-4}
 & & Traffic Steering &\cite{Lacava_Programmable_2023}\\
\cline{3-4}
 & & Unmanned aerial vehicle (UAV) trajectory optimization & \cite{Mohammadi_Trajectory_2023}\\ 
 
\cline{2-4}
 & SARSA & RRM & \cite{Mungari_RL_Approach_2021} \cite{Nagib_Accelerating_2023}\\
 
\cline{2-4}
 & REINFORCE & Radio Resource Allocation & \cite{sharara_policy-gradient-based_2022}\\ 
 
\cline{2-4}
 & \multirow{4}{*}{DQN} & Radio Resource Allocation & \cite{Zhang_Team_2022}\\
\cline{3-4}
 & & UAV trajectory optimization & \cite{Mohammadi_Trajectory_2023}\\
\cline{3-4}
 & & CU-DU placement & \cite{Joda_DeepReinforcement_2022} \cite{Joda_UE_2023}\\ 
\cline{3-4}
 & & Controlling cell activation and deactivation & \cite{bordin_design_2025}\\
 
\cline{2-4}
 & DQN-MARL & Capacity sharing & \cite{Vila_Capacity_2022}\\
 
\cline{2-4}
 & HRL & Traffic Steering; Cell Sleeping; and Beamforming & \cite{Habib_Intent_2023} \cite{habib_machine_2025}\\
 
\cline{2-4}
 & \multirow{2}{*}{D3QN} & BSs' functional splitting & \cite{Murti_DeepReinforcement_2024}\\
\cline{3-4}
 & & Configure the transmission parameters and resources & \cite{Huang_Universal_2022}\\ 
 
\cline{2-4}
 & \multirow{2}{*}{MADRL} & Radio Resource Allocation & \cite{iturria-rivera_multi-agent_2022}\cite{Alsenwi_Coexistence_2022}\cite{Betalo_MultiAgent_2023}\\ 
\cline{3-4}
 & & Power Allocation & \cite{giannopoulos_supporting_2022}\\
 
\cline{2-4}
 & \multirow{2}{*}{Policy Gradient} & VNF scaling and placement & \cite{Amiri_DeepReinforcement_2024}\\
\cline{3-4}
 & & Radio Resource Allocation & \cite{sharara_policy-gradient-based_2022}\\
\cline{2-4}
 & Policy Iteration & Beam Management & \cite{Hoffman_Beam_2023}\\
 
\cline{2-4}
 & \multirow{2}{*}{Actor-Critic} & Radio Resource Allocation & \cite{Hammami_On_Policy_2022}\cite{Kouchaki_Actor_2022}\\
\cline{3-4}
 & & Elastic O-RAN slicing & \cite{abedin_elastic_2022}\\
 
\cline{2-4}
 & \multirow{2}{*}{A2C} & Radio Resource Allocation & \cite{Hammami_Multi_2022}\\
\cline{3-4} 
 & & VNF allocation & \cite{Amiri_Energy_2023}\\
 
\cline{2-4}
 & \multirow{2}{*}{MAB} & Radio Resource Allocation & \cite{Lai_Intelligent_2023}\\ 
\cline{3-4}
 & & VBs scheduling & \cite{Kalntis_BaseStation_2022}\\
\cline{2-4}
 & Neural MCTS & RU-DU resource assignment & \cite{wang_selfplay_2022}\\
\cline{2-4}
 & Optimal Orchestration Policy & Resource Allocation & \cite{ho_energy_2024}\\
\cline{2-4}
 & Parallel Hierarchical DRL & Resource Allocation & \cite{10820044}\\
\cline{2-4}
 & TD3-TS & RRM & \cite{sohaib_optimizing_2025}\\
\cline{2-4}
 & RL-MAML & Resource Optimization & \cite{lotfi_meta_2025}\\ \hline

\multicolumn{1}{|c|}{\textbf{\multirow{10}{*}{FL}}} & Federated Averaging(FedAvg) & MAC Scheduling & \cite{Arslan_Dynamic_2023}\\
\cline{2-4}
 & FRL & Transmission power selection & \cite{Li_Transmit_2022}\\
\cline{2-4}
 & Federated DRL & Multiple xApps coordination & \cite{zhang_federated_2022}\\
\cline{2-4}
 &  \multirow{2}{*}{F-DQN} &  VNF splitting & \cite{amiri_edge-ai_2024}\\
 \cline{3-4}
 & & Offloading and fronthaul routing & \cite{Ndikumana_Federated_2023}\\  

\cline{2-4}
 & F-DRL with DDQN & Radio Resource Allocation & \cite{Rezazadeh_Specialization_2023}\\
\cline{2-4}
 & Federated Meta Learning & Traffic Steering & \cite{erdol_federated_2022}\\
\cline{2-4}
 & F-MARL & Jamming Attack Detection & \cite{houda_federated_2024}\\

\cline{2-4}
 & \multirow{2}{*}{HFL} & Resource Allocation and Scheduling & \cite{wang_hierarchical_2024}\\  
\cline{3-4}
 & & UE Handover & \cite{singh_user_2025}\\  
\cline{2-4}
 & FL-DP-SMC & Enhancing Data Privacy & \cite{yasin_differential_2025}\\  
\cline{2-4}
 & P2P-FL & Cyberattacks Detection & \cite{alalyan_secure_2025}\\  
\cline{2-4}
 & Federated GrINet (FGrINet) & Channel Estimation & \cite{norouzi_decentralized_2025}\\  \hline
\end{tabular}
\end{table*}
}

{\color{black}
\subsection{ML in O-RAN: Advantages, Constraints,  and a Unified Taxonomy} %,
The integration of ML into O-RAN highlights its growing importance, offering notable strengths while also introducing several constraints. These advantages and limitations are summarized below.

\subsubsection{Advantages of ML Utilization in O-RAN}
    \begin{itemize}
         \item \textbf{Enhanced Resource Optimization:} AI/ML can optimize system performance based on prediction accuracy, making it highly effective for radio and spectrum resource management, including increasing throughput and reducing latency \cite{Agarwal_2023}.

    \item \textbf{Anomaly and Threat Detection:} The capability of AI/ML to analyze complex data patterns enables early detection of anomalies and cyberattacks \cite{Başaran_Deep_2023,huang_developing_2023}. Early detection helps maintain network stability through prompt and informed responses.

    \item \textbf{Service Personalization and QoE Improvement:} AI/ML’s adaptive properties allow services to be dynamically tailored to user preferences and behavior, thereby enhancing the Quality of Experience (QoE). For instance, visual and gaming applications benefit from intelligent bandwidth adjustments based on user demand \cite{kougioumtzidis_qoe_2024}.

    \item \textbf{Autonomous Network Operation:} The intelligent and self-learning nature of AI/ML enables O-RAN to operate with higher autonomy, improving management efficiency through faster decision-making and reducing the need for manual intervention \cite{hamdan_recent_2023}.

    \end{itemize}
        
\subsubsection{Constraints of ML Utilization in O-RAN}
    \begin{itemize}
        \item \textbf{Vulnerability to adversarial attacks:} ML models can be targeted by adversarial attacks, which pose a significant risk to O-RAN. Such attacks can manipulate ML algorithms, leading to inaccurate outputs and undermining network integrity, particularly compromising the effectiveness of security defense systems \cite{sapavath2023experimental}.
        
         \item \textbf{Increased computational and energy demands, and model complexity:} Deploying ML in O-RAN introduces significant computational and energy requirements, increasing operational costs \cite{filali_multi-access_2020,latif_energy_2024}. Training and inference add overhead beyond standard O-RAN operations, and achieving higher model performance often requires more sophisticated and resource-intensive architectures. This escalates the processing burden on RIC entities, creating a trade-off between algorithmic accuracy and computational efficiency \cite{makhlouf_optimized_2023,Rastogi_Intelligent_2023,Ntassah_xApp_2023}.

        \item \textbf{Limited labeled data and bias:} Effective ML model training typically relies on large volumes of labeled data. In the dynamic and heterogeneous O-RAN environment, labeled data may be scarce or incomplete, leading to biased models and degraded performance. Mitigating these effects requires adaptive, robust, and semi-supervised learning techniques capable of handling sparse or evolving datasets \cite{Başaran_Deep_2023}.

         \item \textbf{Sensitivity to hyperparameter tuning:} RL, widely applied in O-RAN, is highly sensitive to hyperparameter selection. Since RL learns through trial-and-error interactions, minor parameter misconfigurations, such as the discount factor, learning rate, or policy update frequency, can propagate and substantially affect performance. Precise tuning is therefore essential to achieve stable and optimal outcomes\cite{Pandey_DeepReinforcement_2023}.

         \item \textbf{Communication overhead:} In FL, raw data remains local, and learning depends on frequent exchanges of model parameters. In complex, heterogeneous O-RAN environments, this can generate substantial communication overhead. Techniques such as partial parameter aggregation or selective update sharing can mitigate this overhead while preserving model accuracy and performance \cite{cao_federated_2021}.

 \end{itemize}

Building on the discussion of ML’s benefits and constraints in O-RAN, we propose a taxonomy that systematically classifies ML usage within the architecture, as illustrated in Fig. \ref{fig:taxonomy}. This taxonomy positions ML as a central intelligent component in O-RAN, supporting three key objectives: service quality enhancement, communication quality enhancement, and security quality enhancement.
In the \textit{service quality} category, the primary challenge is resource allocation, with representative use cases including resource allocation optimization and scheduling optimization. These tasks predominantly leverage RL, DRL, FL, and hybrid FDRL techniques due to their adaptability and ability to handle dynamic network conditions.
For \textit{communication quality}, spectrum management is the central challenge, with use cases such as spectrum sharing and allocation optimization. RL and DL techniques are most frequently applied here, providing intelligent and adaptive solutions for dynamic spectrum environments. Within the \textit{security quality} category, use cases cover attack detection, anomaly detection, and traffic prediction. This domain utilizes a broad spectrum of ML techniques—including SL, DL, UL, RL, and FL—reflecting the diversity of security threats and the need for flexible, data-driven defense strategies.

 \begin{figure}[h!]
\centering
\includegraphics[height=4.0cm, width=8.5cm]{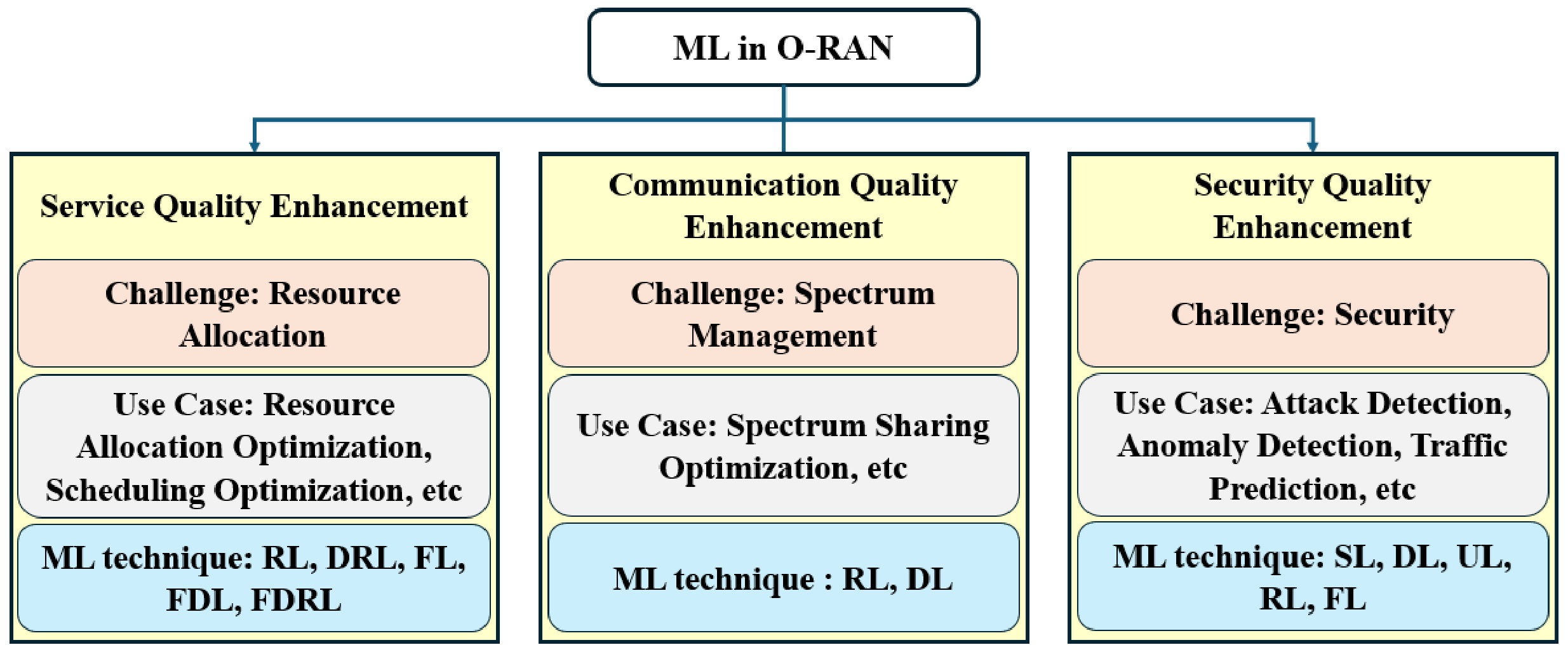}
\caption{{\color{black} A taxonomy of ML in O-RAN.}}
\label{fig:taxonomy}
\end{figure}

Overall, while AI and ML offer substantial potential to optimize performance, enhance efficiency, and strengthen security in O-RAN, their integration must be carefully managed. A balanced approach that considers computational constraints, data availability, security vulnerabilities, and model complexity is critical to ensuring the long-term reliability, resilience, and effectiveness of ML-enabled O-RAN networks.

\subsection{Pre-Deployment Testing of AI/ML Models in the RIC}
The integration of  ML algorithms in O-RAN cannot occur directly in the RIC, as it requires a staged approach involving validation, adaptation, and system integration. First, a simulation environment is needed to safely and repeatedly test ML algorithms within the O-RAN context, ensuring their functionality and performance. Next, an emulation environment is required to align the ML code with actual O-RAN interfaces and protocols, which represents an essential step as transitioning from a purely virtual setup to real-world operation often introduces practical challenges. Finally, before deployment, the ML module must be integrated into the full O-RAN system, ensuring proper communication among all components to ensure smooth and accurate algorithmic operation \cite{bonati_5g-ct_2025,herrera_tutorial_2025}.

\subsection{Lessons Learned}
\begin{itemize}
\item \textbf{Suitability of ML approaches:} Different ML paradigms are best suited for specific O-RAN tasks. SL excels in traffic prediction and anomaly detection, UL is effective for clustering and handover management, and RL is particularly well-suited for adaptive resource allocation and dynamic optimization. RL’s interaction-based learning makes it highly effective in O-RAN’s rapidly changing environments, often complemented with FL to support distributed, privacy-preserving deployments.

\item \textbf{Primary objectives of ML in O-RAN:} ML enhances service quality, communication efficiency, and network security by enabling near-real-time optimization, intelligent RAN slicing, and adaptive decision-making through the RIC framework. These capabilities collectively improve user experience, resource utilization, and operational resilience.

\item \textbf{Key challenges and constraints:} Despite its potential, ML integration faces significant hurdles, including high computational and energy demands, scarcity of labeled data, sensitivity to hyperparameter tuning, and vulnerability to adversarial attacks. Overcoming these challenges requires the development of lightweight, interpretable, safe, and energy-efficient ML models tailored for O-RAN’s distributed architecture.

\item \textbf{Testing and validation:} Pre-deployment testing of ML modules in simulated O-RAN environments is essential to ensure interoperability, stability, and compliance with SLAs. Rigorous validation helps prevent performance degradation and ensures safe and effective deployment in real-world networks.

\end{itemize}

}

{\color{black}
\section{ML for Tackling Challenges in O-RAN }
{\color{black}
As a core intelligent component enabling data-driven decision-making, ML has emerged as a transformative technology within O-RAN. This section analyzes how different ML techniques are applied to tackle critical challenges aligned with the previously introduced taxonomy, including enhancing spectrum management, optimizing resource scheduling and allocation, and reinforcing network security. By leveraging ML’s adaptive and predictive capabilities, O-RAN can achieve more efficient, reliable, and resilient operations across its distributed and dynamic architecture.}

\subsection{Spectrum management}
Spectrum management has evolved from the static allocation of set frequency bands to designated entities in the previous wireless network generations.In 5G and beyond and within the O-RAN architecture, third-party applications, xApps, and rApps, deployed in the RICs, have enabled mobile operators to directly incorporate AI/ML algorithms into the network.  These solutions enable advanced, near real-time, and long-term dynamic spectrum optimization and resource management \cite{sapavath2023experimental, Polese_ColoRAN_2023, 10.1109/tnsm.2024.3373606}

Various spectrum allocation strategies, such as cognitive radio (CR) technologies \cite{9237455,9408651,9612017,9838746,9926102}, dynamic spectrum access (DSA), and spectrum sharing (SS) have been explored to satisfy dynamic and diverse service requirements in the context of O-RAN \cite{sharma_overview_2025, song_deep_2021, 10.1109/access.2019.2922702, Khan_2020, 10.21203/rs.3.rs-1364812/v1, Kaur_Kumar_2022, Nair_2023}. For instance, CR is known for its adaptive and intelligent ability to automatically detect idle channels in the wireless spectrum and adjust transmission parameters to improve spectrum allocation through efficient frequency band utilization \cite{9500621,10278964,awin_blind_2019,10466378,10464644,9348134} thanks to AI/ML techniques. This is achieved by allowing secondary users (SUs)  to dynamically share underutilized licensed spectrum bands without causing significant interference to primary users (PUs), consequently enhancing the spectral efficiency \cite{awin_technical_2019}. This can be effectively adopted in O-RAN, as its architecture is designed to support a vast number of devices and can integrate ML algorithms to enable efficient and scalable spectrum management  \cite{Ahmed_Chen_2021,Nair_2023, Kaur_Kumar_2022}.

Papers as \cite{Gopal_2024, Asad_Otoum_2024,song_deep_2021, Gopal_Griffith_2024} proposed different O-RAN-compatible AI/ML strategies like gradient boost trees, LSTMs, and RNNs to attain efficient dynamic spectrum management. These various works introduced solutions from building intelligent radio resource demand prediction to proposing data-driven spectrum management schemes and xApps of RL models capable of efficiently, autonomously, and dynamically managing the spectrum utilization by learning network demand patterns and using them to allocate resources. Inspired by a DQN, the authors in \cite{song_deep_2021} suggested a DRL framework for dynamic spectrum access in heterogeneous networks.  By allowing users to make individual decisions on spectrum access and power allocation without depending on centralized control or full channel state information (CSI), their methodology allows distributed spectrum management.  Optimizing these parameters reduces interference and delay and enhances data rates. Moreover,  in order to accomplish real DSA, \cite{8935684} and \cite{6923334} have emphasized the need to detect and categorize interference sources, whether they are from users within the network, those from outside, or even jammers in a wireless environment, using AI/ML algorithms. \cite{8935684} specifically presented a DL signal modulation model as a classification solution in realistic conditions and considering multiple scenarios. The O-RAN architecture with the two RICs has definitely paved the highway to AI/ML algorithms application in 5G and beyond RANs. Developed models will be running as near-real-time xApps, since the spectrum access is subject to changing environments, interference management, changes in the radio state,  and other external conditions.

\subsection{Resource allocation}
The ambition with the O-RAN to make the B5G RANs intelligent and dynamic in real-time naturally brings in the challenge of efficient resource allocation \cite{azariah_survey_2024, Brik_DeepLearning_2022, 9798822}.  The most critical resource allocation tasks to tackle include radio resources, computation resources, and power control. Given the importance of resource availability and network robustness to failures for different 5G services, in particular, real-time applications, AI/ML techniques stand out to offer effective solutions to address these challenges and improve the performance and adaptability of O-RAN networks \cite{9120231, Brik_DeepLearning_2022, brik_explainable_2025}.

\subsubsection{Radio resource allocation}
Radio resource allocation in O-RAN is challenging because of the limited radio resources to be shared among diverse users with various demands and service requirements under a near-real-time constraint. Moreover, the RAN slicing concept underlying O-RAN further complicates this challenge, as it involves partitioning network resources to meet diverse and service-specific requirements \cite{10921752, 10433640}. The traditional resource allocation methods, such as closed-loop control systems\cite{Ndikumana_2023}, multiparameter optimization methods \cite{Lira_2023}, bio-inspired heuristics \cite{Markovic_2017, Wright_2014}, or QoE-driven optimization algorithms\cite{Agarwal_2023}, often fail to efficiently manage these slices, particularly under dynamic network conditions \cite{giannopoulos_supporting_2022, azimi_applications_2022, chang_toward_2023}. This is because they are deterministic, rule-based, and they focus on one aspect of the network at a time. In this context, a novel approach is emerging through the use of quantum computing, leveraging the advantage of quantum parallelism \cite{11059714,kalfon2024successive,mlika2023user,10821473,cherkaoui2023quantum,10811384}. However,  in practice, quantum algorithms are limited by the hardware deficiencies, including the number of qubits, the noise, and the small coherence time, rendering them non-scalable on NISQ devices \cite{10921752}. 

The difficulty in O-RAN lies in ensuring that each slice meets its QoS requirements while maximizing overall resource utilization. Given this,  ML models, particularly RL, have already proven to be effective in optimizing radio resource allocation.
 \cite{Tripathi_2023} presented a real-life testbed with an end-to-end 5G-based O-RAN deployment that leverages AI/ML models for intelligent radio resource allocation, deployed in both non RT RIC and near RT RIC for near real-time and long-term resource management. DRL algorithms have been extensively examined for their adaptability to diverse environments by dynamically allocating resources according to real-time network conditions and user behavior \cite{10667952, 10772596, 10820044, wang_energy-efficient_2024, 10741435,li_deep_2018, abiko_flexible_2020}, making them convenient to use in O-RAN deployments. This assists the system in dynamically learning the most effective strategies for allocating limited resources among competing requirements within the framework of resource allocation, thereby adapting to evolving network conditions and usage patterns over time.   Ultimately, this approach leads to a substantial increase in the efficiency of resource allocation. In contrast to DRL, multiple works \cite{khodapanah_framework_2020, 10920786, 10625184} suggested the integration of intelligence in O-RAN through AI/ML resource management frameworks to predict network behavior and allocate resources to fulfill the service level specifications. By leveraging historical performance data, they can provide insights into future resource needs to proactively make needed adjustments. This enhances the service delivery and user satisfaction. Furthermore, by deploying ML models in the O-RAN architecture, operators can achieve better performance and adaptability in managing radio resources and in addressing the consequential complexities due to the disaggregated nature.

\subsubsection{Computation Resource allocation}
Computation resource allocation in O-RAN environments presents important difficulties, particularly because of the need to process the demands of various applications in cloud and edge computing scenarios. It is critically challenging to efficiently offload computational tasks to cloud resources while minimizing latency and maximizing throughput \cite{11178232, rodoshi_survey_2021}. The computation resource management becomes more and more complex when the number of users and the variability of their computational demands increase. ML techniques and algorithms are excellent tools to settle these challenges. For instance, the authors in  \cite{ma_joint_2020} proposed using ML algorithms to predict computational demands based on user behavior and application requirements. By anticipating peak demand periods, the network can allocate resources dynamically, ensuring that computational capabilities are aligned with user needs. On the other hand, \cite{azimi_applications_2022}  highlighted the potential of ML techniques in resource management for RAN slicing, indicating that adaptive algorithms can optimize computation resource allocation based on real-time traffic patterns. The adaptability of AI/ML algorithms, particularly the DRL models, is crucial for satisfying the processing and latency requirements of a wide range of applications. For example, the authors in \cite{Hammami_On_Policy_2022} have proposed two DRL models to solve the O-DU computational resource allocation for latency-sensitive tasks and latency-tolerant tasks in an O-RAN network. This is to showcase the advantage of RL   over greedy and traditional methods in the context of diverse QoS requirements. Real-time video transmission and latency-tolerant application tasks are simulated within a slicing-based O-RAN system.  Under limitations that guarantee latency remains below a certain level and prevent exceeding resource capacity, computing resources (CPU cores) are allocated from virtualized O-DUs to service these tasks in each time window. Minimizing the total power consumption of the O-DUs is the objective of this scenario. While this optimization problem can be formulated as a mixed-integer programming (MIP) model and solved using classical solvers, such methods suffer from poor scalability in large and dynamic environments. Therefore, the authors made use of DRL techniques and modeled the CPU cores allocation process as a Markov decision process. Hence, the agent's environment consists of a finite state space (all users' demands and the O-DU resources utilization state at a given time slot), a finite action space (O-DU CPU cores allocation to users), and the reward function (the negative of the power consumption based on the action taken). The authors utilized solely power consumption in modeling the reward function, which is excessively restricted.  By adjusting the reward function to incorporate penalties for excessive power consumption, high latency, and violations of the established power and latency thresholds, we have achieved more stable convergence for the two DRL models suggested in \cite{Hammami_On_Policy_2022}, as shown in Fig. \ref{fig:reward1}.  The actor-critic with experience replay (ACER) and PPO models have been considered for the simulation. These methods are both model-free algorithms as they do not involve environment modeling or next state prediction\cite{9275964}, yet they determine the optimal policy by estimating the value function for each state-action pair. It is worth mentioning that the choice of these model-free algorithms is suitable for 5G and beyond O-RAN networks, as the network environment and dynamics can vary significantly even in the same physical area of the networks. 
\begin{figure}[htbp]
    \centering
    \includegraphics[width=0.95\columnwidth]{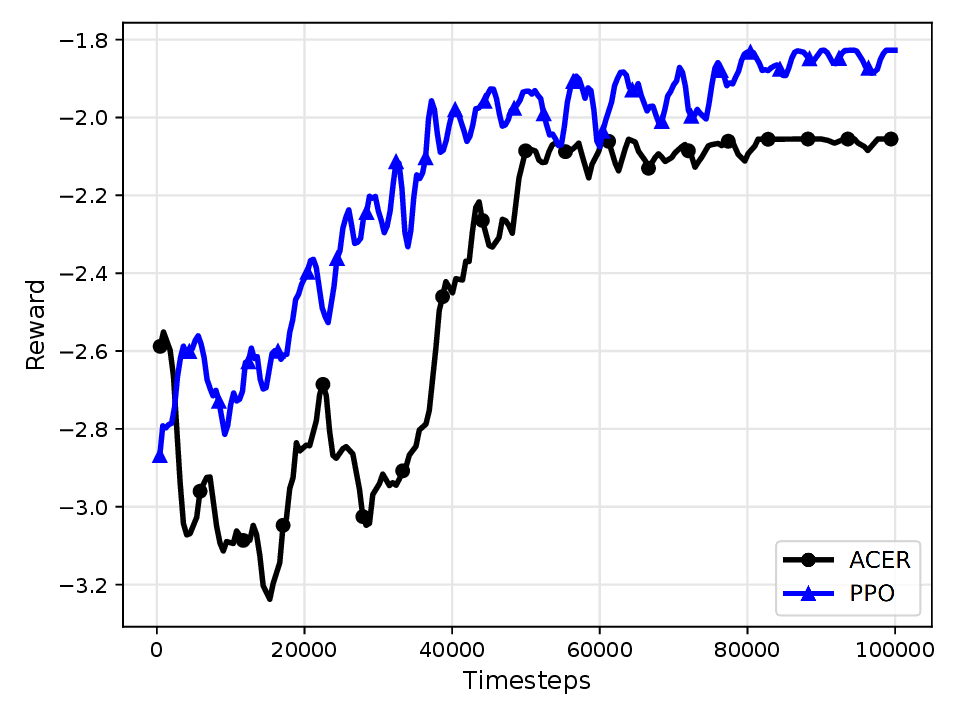}
    \caption{Reward function over the simulation time-steps for ACER and PPO models for O-DU computation resource allocation }
    \label{fig:reward1}
\end{figure}

\begin{figure}[htbp]
    \centering
    \includegraphics[width=0.95\columnwidth]{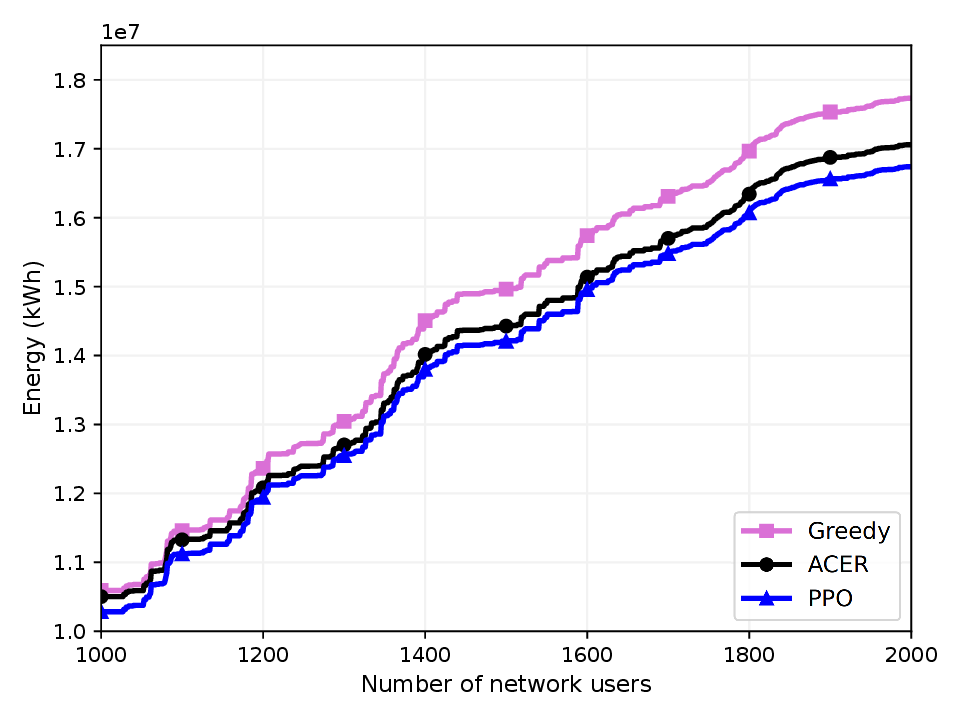}
    \caption{The power consumption at the O-DU as a function of the number of incoming requests}
    \label{fig:power}
\end{figure}

\begin{figure}[htbp]
    \centering
    \includegraphics[width=0.95\columnwidth]{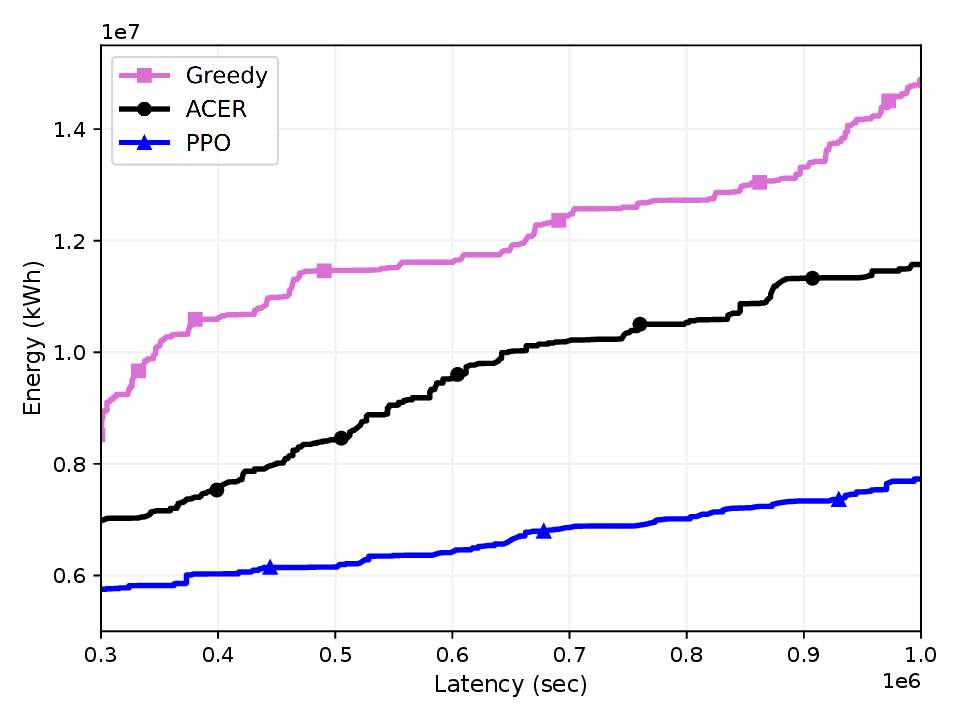}
    \caption{The power consumption at the O-DU over time}
    \label{fig:latency}
\end{figure}

The performance of the proposed enhanced DRL-based resource allocation framework is evaluated through simulations, as illustrated in Fig. \ref{fig:reward1}, \ref{fig:power}, \ref{fig:latency}.  Fig. \ref{fig:reward1} shows the reward function versus the time steps for both ACER and PPO. We notice that both techniques converge toward higher rewards. However, ACER achieves faster and more stable convergence than PPO. This could be explained by ACER’s experience replay mechanism, which efficiently reuses past transitions to improve policy updates. PPO, on the other hand, struggles with the multi-dimensional reward structure (power, latency, and thresholds), leading to slower and less stable adaptation.

Figs. \ref{fig:power} and \ref{fig:latency} show the energy consumption (kWh) at the O-DU over the number of network users (ranging from 1,000 to 2,000) and the energy consumption (kWh) at the O-DU over elapsed time, respectively. In both figures, we compare the DRL techniques with the Greedy policy. The Greedy algorithm prioritizes simplicity by selecting the server with the lowest CPU utilization for each task. While it is simple to implement and computationally lightweight, it lacks intelligence and operates blindly, focusing only on minimizing immediate resource usage without considering latency constraints or future demand fluctuations. This naturally leads to suboptimal performance in dynamic O-RAN environments, where the energy consumption significantly increases under high user loads and with time.
In contrast, ACER (off-policy) and PPO (on-policy) employ DRL to balance long-term trade-offs. ACER uses experience replay to reuse past transitions, improving sample efficiency. However, its reliance on historical data makes it sensitive to hyperparameter tuning and less adaptable to network changes. Nevertheless, PPO leverages a clipped objective function to stabilize policy updates, ensuring gradual adaptation to dynamic conditions. This enables PPO to constantly maintain lower energy consumption, as witnessed in Fig. \ref{fig:power} and \ref{fig:latency}.

\subsubsection{Power control}
Power control is as important as the radio and computation resources management in O-RAN environments to maintain energy efficiency while ensuring reliable communication. The need to balance power consumption with the QoS is critical, especially in dense urban environments where interference can significantly impact performance \cite{10.1109/access.2019.2960490}. AI/ML algorithms had already been proposed for previous generations of RAN \cite{luo_power_2020, rodoshi_survey_2021}, as naive approaches--such as those discussed in the previous subsection--are overly simplistic and unsuitable for long-term optimization. Meanwhile, classical optimal solutions lack scalability and become ineffective in 5G and beyond networks characterized by massive device connectivity and diverse service requirements. For instance,  \cite{10.3390/s22135029}  clearly demonstrated how greatly energy efficiency may be improved by optimizing computational processes.  The authors used actor-critic learning in a DRL framework to implement energy-aware dynamic selection of O-DUs inside an O-RAN architecture.  Their results confirm the efficiency of AI/ML methods in jointly optimizing resource allocation and energy consumption.
The ability of O-RAN systems to respond dynamically and often in real-time requires large amounts of data collection, storage, processing, and constant monitoring.  These operations result in a substantial increase in energy consumption, which is the primary resource that maintains system functionality, as a result of the increased computational workload. The scalability and efficiency of O-RAN are severely challenged by this direct relationship between real-time responsiveness and energy demand. Several previous works \cite{luo_power_2020, shen_hybrid_2022} attempted to address this vital aspect of O-RAN by using supervised and RL to specifically enhance power control. Predictive models that optimize power allocation in near real time are typically developed by utilizing historical power usage patterns and user demands. This is a prevalent approach among the proposed supervised algorithms. For instance, the authors of \cite{Liang_Al_2024} have successfully implemented intelligent xApps in an O-RAN network, resulting in a $50\%$ reduction in power consumption.  Instead, \cite{Chen_2024} developed a statistical learning approach that is AI-based. This approach integrates the detection of O-RAN abnormalities at the BS with an effective power control mechanism. In conclusion, the incorporation of AI/ML into O-RAN systems is a successful strategy that will lead to the development of effective power management techniques.  AI/ML techniques are a potent way to optimize power distribution across various O-RAN components and improve the overall system efficiency, whether by addressing joint optimization problems, leveraging real-time data analytics, historical data, and energy consumption patterns, or developing specific energy-oriented solutions.

}
{\color{black}
\subsection{Security}
O-RAN security research is strongly driven by the huge benefits and innovations O-RAN brings to the cellular network industry regarding its observability, reconfigurability, and cost-efficiency \cite{groen2023implementing,liyanage2023open}. However, the adoption of O-RAN brings new security challenges due to the technology being cloud-based, multi-vendor, and open in nature, thus increasing the attack surface and exposing the network to cyber-attacks \cite{groen2023implementing}. Hence, the security analysis related to O-RAN systems is necessary for exposing any vulnerabilities or threats to the integrity and confidentiality of the network operations \cite{mimran2022evaluating,liyanage2023open}. By exploring the security aspects of O-RAN, researchers are working toward investigating the state of security within O-RAN in order to discover threats and offer relevant solutions \cite{groen2023implementing}.

{\color{black}
A recent security assessment \cite{citation-key} provides a more detailed view of these vulnerabilities and their relative significance within O-RAN deployments. Although most threats in O-RAN environments resemble those found in traditional RAN systems, a small but critical subset, approximately 4\% of the total identified threats, is unique to O-RAN. The majority of these threats are classified as high risk and are concentrated around sensitive interfaces and components, including Non-RT RIC, rApp, A1, E2, and R1. In addition, O-Cloud infrastructure accounts for approximately 18\% of the highest-risk threats, reflecting its strategic significance as a central and potentially vulnerable element within the architecture. These observations emphasize the need for the adoption of zero-trust security architectures, stronger access control mechanisms, and continuous monitoring of critical operational layers.}

  O-RAN will feature open interfaces and disaggregated components, rendering it flexible and interoperable. However, this approach also increases the potential number of vulnerabilities \cite{chen2023brief}. Given this, ML has recently been recognized as a powerful tool in enhancing the security of O-RAN. ML specializes in addressing security challenges, with advanced capabilities in threat detection, prediction, and response. The application of ML in cybersecurity frameworks automates decision-making processes, ensuring rapid responses to threats and establishing a robust defense against growing cyber risks \cite{okoli2024machine}.  In the context of O-RAN, ML is crucial for network automation, addressing the complexities associated with managing multivendor and interoperable solutions, while enhancing the overall security posture of the network  \cite{hamdan_recent_2023}. For instance, efficient supervised approaches include decision trees and support vector machines (SVM) that detect well-known patterns of attacks against network traffic \cite{mirlashari2023machine}. In the event of limited available labeled data, unsupervised techniques such as clustering and anomaly detection identify novel intrusion patterns by highlighting deviations from normal behavior \cite{anurag2024robotic}. Moreover, ML enables threat intelligence in predictive analytics by analyzing historical data trends to predict future attacks, allowing defense strategies to be proactive \cite{machhindra2023enhancing}. For example, RL develops automated response systems that adaptively implement the security policies based on the evolving threats and make immense improvements in real-time threat mitigation \cite{anurag2024robotic}. 

ML strengthens mechanisms of authentication and access control through the analysis of biometric data and improvement of role-based access control to realize secure access to network equipment resources \cite{anurag2024robotic}. Specifically, ML methods, including a novel open-set detection approach based on CNN and long short-term memory (LSTM) models, have been proposed to identify unauthorized devices from RF signal patterns at the air interface and prevent unauthorized network access. \cite{puppo2024extraction}. Likewise, DL models can provide spectrum access techniques that guarantee data privacy using encryption methods. This is illustrated through a shuffling-based learnable encryption technique combined with a Vision Transformer (ViT) model that, despite operating on encrypted data, showed vast improvements in accuracy and F1-Score \cite{gajjar2024preserving}. Furthermore, ML models, including CNNs and DNNs, have been used in xApps by the near-real-time RIC for countering such adversarial attacks. Techniques such as distillation, developed to improve the resiliency of these models, maintain remarkable accuracy even under attack conditions \cite{chiejina2024system}. Particularly, distillation is a technique in which a smaller, simpler model (student) is trained to replicate the behavior of a larger, more complex model (teacher), often improving the model's robustness and generalization, especially under adversarial conditions  \cite {sun2024diredi}

This section presents a comprehensive review of ML applications in O-RAN, demonstrating that ML methods can effectively enhance security measures to address the challenges posed by the intelligent and open nature of O-RAN ecosystems. We will highlight how ML can revolutionize O-RAN security by providing dynamic, intelligent, and adaptive solutions to emerging security challenges. Within the following subsections, we will investigate the diverse security challenges related to O-RAN and evaluate existing studies and applications of ML techniques that have addressed these concerns.

\subsubsection{Security Challenges of Open Interfaces}
Recent research has been focused on how AI/ML could secure the open interfaces of O-RAN. The disaggregated nature of O-RAN, which promotes openness and interoperability, introduces security vulnerabilities, particularly in its open interfaces such as the E2 interface and Open Fronthaul. Due to the interoperability between multiple vendors, these interfaces are more exposed to cyber threats, including eavesdropping, man-in-the-middle attacks, and unauthorized access. Without stringent security mechanisms, attackers can exploit vulnerabilities in these interfaces to disrupt communication, inject malicious traffic, or compromise sensitive network data.

Encryption protocols have been studied to mitigate these vulnerabilities, revealing that while they enhance security, they also introduce latency and reduce throughput, necessitating a careful cost-benefit analysis \cite{groen2024securing}. Similarly, the reliance on virtualization and software functionality in O-RAN expands the threat surface, making it susceptible to hacking and data theft, particularly in the context of hyperconnected 6G networks \cite{9881863}. These challenges highlight the need for advanced AI-driven security mechanisms to dynamically adapt to emerging threats. 

To address these challenges, researchers have proposed various AI/ML-driven security solutions, such as SL algorithms for cell traffic prediction and DRL for energy-efficiency maximization, which are implemented through xApps on the RIC \cite{giannopoulos_supporting_2022}. For example, in \cite{yungaicela2024misconfiguration}, it is shown that the open nature of O-RAN and the support of heterogeneous systems increase the misconfiguration risk, which can be mitigated by several AI/ML-based solutions identifying and solving the conflicting policies between xApps. The approaches used include anomaly detection, which leverages AI/ML algorithms to monitor KPIs and detect deviations from normal behavior that may indicate misconfigurations. Correlation analysis is also employed to identify relationships between different xApps and their impact on system performance, determining which xApps may be conflicting. Additionally, active monitoring techniques are utilized where AI/ML sends synthetic service requests or probe packets to interact with the system and uncover misconfigurations. Furthermore, conflict resolution algorithms are applied to mitigate the conflicting objectives of independently operating xApps once detected. Finally, the development of a unified detection framework that integrates various AI/ML techniques is advocated to enhance overall detection and resolution of misconfiguration issues in the O-RAN environment. In \cite{kouchaki2023openai}, federated RL  (FRL) is proposed to make a shift from presently centralized approaches towards a distributed realization of real-time applications in order to gain both security and efficiency for O-RAN scenarios. By decentralizing learning and processing, federated RL minimizes the exposure of sensitive data to centralized servers, reducing the risk of data breaches while also optimizing decision-making latency. This approach aligns well with the open interfaces in O-RAN, enabling secure and efficient coordination among diverse network entities without relying on a single point of control. 

The O-RAN Alliance has made invaluable contributions to defining AI/ML workflows and specifications that enable secure and efficient operations. The implementation of these workflows through open-source software such as Acumos and  Open Network Automation Platform (ONAP) further supports this effort \cite{lee_hosting_2020}. In \cite{lee_hosting_2020}, the authors designed an AI/ML workflow according to working group 2 (WG2) AI/ML specifications and realized it using open source software from the O-RAN SC, Acumos, and ONAP. The Acumos Framework was used to generate and package ML models to be deployed and executed in the O-RAN  Intelligence Controller (RIC), while components of Open Network Automation Platform (ONAP) provided monitoring and arbitration needed to operate the workflow.
The AI/ML models deployed via this workflow can enhance security by identifying anomalous behavior and potential threats within the network. The continuous monitoring capabilities offered by ONAP further bolster security by allowing for real-time assessments of network performance and security posture. Overall, the use of standardized open interfaces within this workflow not only ensures compatibility across various components but also allows developers to create more secure and flexible systems, thereby facilitating better risk management and compliance with best practices. 

Autonomous fault management systems, such as the open Fault Management (openFM) framework, leverage AI/ML to predict and manage faults, thereby enhancing the reliability and security of O-RAN networks \cite{mukherjee_open_2023}. For example, in \cite{cheng2024real}, they demonstrated that using ML for real-time inference enhances O-RAN security by enabling more efficient and accurate processing of CSI feedback, which can help in detecting anomalies and potential security threats in the network. The paper specifically employs an autoencoder-based model for CSI compression to facilitate this real-time inference. This approach allows for improved scalability and adaptability in security measures within the O-RAN framework.

Despite these advancements, the open and programmable nature of O-RAN necessitates a cautious approach to security, with ongoing efforts to standardize and implement robust security measures \cite{liyanage2023open}. Collectively, these research efforts underscore the critical role of AI/ML in securing open interfaces in O-RAN, highlighting the need for robust, explainable, and distributed AI/ML solutions to address the unique security challenges posed by the open and disaggregated nature of O-RAN.

\subsubsection{Supply Chain Security}

ML techniques can effectively address security challenges that arise within the O-RAN supply chain, which includes the diverse ecosystem of hardware, software, and service providers responsible for building, integrating, and maintaining O-RAN components. Since O-RAN promotes vendor diversity through open interfaces, its supply chain involves multiple entities, increasing the risk of security threats such as compromised firmware, malicious software updates, or vulnerabilities in third-party components. The integration of ML is essential as the supply chain has inherent threats that can be exploited at various stages, posing significant risks to business continuity. ML techniques, including algorithms like SVM and RF, are utilized to develop threat intelligence systems capable of identifying which nodes within the cyber supply chain are most vulnerable to attacks, thus enhancing the organization's ability to maintain security \cite{yeboah-ofori_cyber_2021}. ML can analyze large datasets to identify abnormal patterns, predict potential vulnerabilities, and enhance threat detection and response times, thereby improving overall supply chain security.

O-RAN, which specifies interfaces that allow equipment from different suppliers to work together, provides network flexibility at reduced cost but also raises new security and privacy issues \cite{liyanage2023open}. The integration of Cyber Threat Intelligence (CTI) with ML techniques has been shown to significantly improve the analysis and prediction of cyber threats targeting supply chain security. This combination enables the systematic identification of vulnerabilities within the supply chain ecosystem and supports organizations in implementing timely, effective control measures to strengthen their overall cybersecurity posture. This ensures resilience against potential attacks while preserving the integrity and operational continuity of their supply chain systems\cite{yeboah-ofori_cyber_2021}.

In order to prevent escalated privilege attacks that have the potential to compromise internal networks, unsupervised ML approaches have been utilized to profile the typical behaviors of privileged users and create risk score functions to identify anomalies \cite{chen2023detecting}. Supply-chain poisoning and identity and access management tampering are two of the unique security concerns that have arisen from the granularization of network services in 5G networks, including O-RAN. For these software-centric architectures, ML models could provide dynamic and reliable security mechanisms that automate effective security measures and improve threat intelligence \cite{afaq2021machine}.

ML-based methods have been proposed to address security challenges in SS systems, which play a crucial role in the operation of O-RAN. Similar to how supply chains thrive on collaboration and resource sharing, SS enables multiple users to access limited frequency bands; however, this sharing introduces various security threats, such as jamming attacks that disrupt communication, eavesdropping that compromises privacy, and issues like Primary User Emulation (PUE) and Spectrum Sensing Data Falsification (SSDF)\cite{wang_when_2022}. ML offers sophisticated tools capable of mitigating these threats by enabling the identification of anomalous user behavior and enhancing the detection of attacks through comprehensive analysis of spectrum sensing data. Thus, the integration of ML techniques not only improves the overall efficiency of SS systems but also bolsters their security framework, ensuring reliable and resilient communication in increasingly complex wireless environments\cite{wang_when_2022}. 

% Deep meta-learning combined with multi-task learning can optimize supply chain operations by leveraging shared domain knowledge, overcoming traditional limitations, and enabling the development of flexible models that enhance security and business performance. This is achieved  by enabling more accurate predictions and faster adaptability to new tasks, even with limited data, which is crucial in dynamic supply chain environments. This adaptability allows companies to respond swiftly to disruptions, enhance decision-making, and ultimately improve customer satisfaction while reducing operational costs\cite{zhu2023overcoming}.

The utilization of datasets, such as the Microsoft Malware Predictions dataset, has demonstrated that algorithms such as Random Forest (RF) and LightGBM are capable of accurately predicting cyber threats, thereby allowing businesses to proactively mitigate supply chain risks \cite{al2022predicting}. Building on this, the effectiveness of ML in risk prediction extends beyond cybersecurity to broader supply chain management. Random Forest, in particular, has shown its versatility in both domains, while more advanced DL models, such as Deep CNN, further enhance predictive capabilities. This is because CNN is capable of accurately predicting risks and handling complex, nonlinear interactions between variables \cite{mittal2023ai}.

Ensuring the security of the supply chain is critical, as vulnerabilities in upstream software components can expose the entire network to cyber threats. The development of tools, such as SPatch, which is based on fine-grained patch analysis and differential symbolic execution, has served in the detection of safe patches to ensure secure software updates that raise the security bar of upstream software in the supply chain \cite{luo2024strengthening}. By strengthening the integrity of software updates, such tools help mitigate risks within the O-RAN supply chain and improve overall network resilience.

Overall, the ML techniques that solve the challenge of supply chain security in O-RAN include predictive analytics, anomaly detection, dynamic security mechanisms, and automated screening. These could improve supply chain security operations and further increase efficiency in this fast-changing landscape of network technology.

\subsubsection{Data confidentiality}

The use of AI/ML to secure O-RAN with respect to data confidentiality is a complex problem that has attracted a lot of interest in recent studies. Data confidentiality is further complicated by the flexibility and interoperability of O-RAN, particularly in the next 6G networks, due to the expanded threat surface from virtualization and software functions. Strong AI/ML-based security solutions are essential as the hyperconnectivity of 6G applications raises concerns about data, location, and identity privacy \cite{9881863}. For example, using DL techniques to improve spectrum access while maintaining data privacy is one well-known approach. Moreover, spectrograms and other sensitive wireless data kept in common databases or multi-stakeholder cloud environments can be secured with encryption methods developed using AI/ML.  \cite{gajjar2024preserving} offered a shuffling-based learnable encryption method integrated with a custom ViT model. When compared to more complex designs, such as ResNet-50 and traditional CNN, this technique significantly improves model accuracy and decreases prediction time. In addition, \cite{groen2024securing} examined the effects of different encryption protocols on throughput and latency, highlighting the importance of encryption in protecting O-RAN interfaces such as the E2 interface and Open Fronthaul. The authors suggested four essential guidelines for building security by design within O-RAN systems. First, sufficient compute resources must be provisioned to ensure that the disaggregated nodes can handle security protocols without negatively impacting performance. Second, the choice of specific protocol implementations and encryption algorithms is crucial, as selecting the right ones can enhance security while minimizing performance overhead. Third, it is important to address Input/Output bottlenecks in both user space and kernel space that could hinder network performance when security measures are applied. Finally, designers should optimize the network Maximum Transmission Unit (MTU) size to facilitate efficient data transmission and avoid delays caused by packet fragmentation. These guidelines aim to help system designers create secure O-RAN architectures while maintaining optimal performance.  In \cite{wen2022fine}, it is shown that the employment of AI/ML-driven security services, such as MobiFlow can offer fine-grained telemetry streams that provide highly detailed and real-time network data tailored for security analysis. These telemetry streams continuously monitor network activity at a granular level, enabling the detection of subtle anomalies, such as unauthorized data access or abnormal traffic patterns. By capturing and analyzing these insights, MobiFlow allows for intelligent security control, enhancing threat detection and response mechanisms. This enables real-time monitoring to mitigate risks, including data theft and malicious transmitters.

Due to the disaggregation of O-RAN and its reliance on open interfaces and AI/ML to enhance RAN operations, security must be carefully managed, as improper management could result in severe privacy concerns \cite{liyanage2023open}. Furthermore, effective data management practices are crucial for safeguarding privacy-sensitive information, as data leakage through communication services remains a significant concern. Proper handling of user data, including identity, location, and personal information, necessitates implementing robust security measures, such as encryption and access control. If these security measures are mismanaged, it could lead to significant privacy concerns \cite{liyanage2023open}. In order to improve data confidentiality, \cite{puppo2024extraction} presented an open-set detection technique for RF data-driven device identification that significantly enhances data management in network security. This approach preprocesses RF signals to filter noise, normalize signal strengths, and handle variations, ensuring reliable input for DL models. By leveraging LSTM networks, the technique extracts unique device fingerprints for accurate real-time differentiation between authorized and unauthorized devices. Effective dataset handling through careful partitioning and cross-validation allows for robust evaluation under open-set conditions. Additionally, the system's real-time processing capabilities enable prompt identification of anomalies or unauthorized devices, crucial for maintaining data integrity and mitigating security threats. Furthermore, the RIC's implementation of AI/ML-assisted algorithms emphasizes the importance of data management security within O-RAN architectures. Techniques like SL and RL facilitate secure, data-driven decision-making while protecting sensitive information. By utilizing network telemetry for real-time data collection, the RIC ensures that data is managed securely throughout its lifecycle. This disaggregated approach not only allows for multi-vendor collaboration but also adheres to strict data protection protocols, fostering a resilient and privacy-conscious network environment \cite{giannopoulos_supporting_2022}. Data confidentiality may be compromised by conflicting policies among xApps, which is why it is important to have strong procedures in place to detect and resolve misconfiguration in O-RAN, especially when AI/ML is being used \cite{yungaicela2024misconfiguration}.  

All these studies together demonstrate how important AI/ML is to O-RAN security, especially when it comes to data privacy in the constantly changing world of next-generation cellular networks.

\subsubsection{Safety of AI} 
  The integration of AI/ML algorithms into RAN management is made possible by the virtualization and network slicing aspects of 5G, which are essential to O-RAN and highlight the necessity of strong security frameworks to preserve data confidentiality \cite{masur_artificial_2022}. That is, small modifications to input data can significantly impact the performance of ML applications, particularly interference classifiers within the near-real-time RIC. These classifiers depend on specific data inputs like spectrograms and KPMs for accurate network interference assessment. Adversarial attacks that manipulate this input data can lead to incorrect classifications, compromising the system's ability to effectively detect interference. This vulnerability is heightened by O-RAN's open architecture, which exposes it to cybersecurity threats that could disrupt AI-driven decision-making. Therefore, there is an urgent need for robust security measures to protect AI components within O-RAN, ensuring their operational integrity against such adversarial attacks \cite{chiejina2024system}. Experimental deployments demonstrating up to 100\% degradation in model accuracy under adversarial conditions indicate that such attacks can cause substantial declines in network performance \cite{chiejina2024system}.
  
{\color{black}
However, it is critical to recognize that this perspective addresses only one dimension of the security challenge. While AI/ML can enhance O-RAN defenses against traditional network threats, the integration of machine learning models introduces a parallel set of vulnerabilities that fundamentally transform the threat landscape. ML models themselves become potential attack vectors, requiring specialized countermeasures across their entire lifecycle, from training through deployment and operational monitoring \cite{chiejina2024system, groen2023implementing, aryal_moving_2023}}
}.
AI/ML applications, deployed as rApps and xApps, serve as critical decision-making engines within O-RAN systems; however, they are vulnerable to threats such as data poisoning and adversarial attacks, which can undermine the integrity and accuracy of their outputs. For instance, attackers may inject misleading data into the training datasets, leading to suboptimal or erroneous decisions regarding network slicing and resource allocation. 

{\color{black} Beyond these traditional evasion attacks, ML models face additional critical vulnerabilities during the training phase. For instance, data poisoning attacks allow adversarial network participants to corrupt training datasets by injecting false measurements, such as exaggerated interference reports or falsified KPIs, causing trained models to learn biased behaviors that persist throughout their deployment lifetime \cite{chiejina2024system}. Model poisoning in federated learning environments, increasingly proposed for collaborative O-RAN intelligence, can compromise global models when malicious participants manipulate their local model parameters during distributed training rounds \cite{aryal_moving_2023}. Furthermore, backdoor attacks can embed hidden triggers into trained models, causing them to behave normally under standard conditions while activating malicious functionality when specific patterns are detected, potentially granting unauthorized network access or degrading service quality \cite{chiejina2024system}. Privacy attacks, including gradient leakage in federated learning scenarios and membership inference attacks, can expose sensitive network data and operational patterns despite privacy-preservation efforts \cite{aryal_moving_2023}. Once deployed, models also face model extraction attacks where adversaries with query access create surrogate models to understand decision boundaries and enable more sophisticated attacks, as well as concept drift where models degrade as network behavior evolves over time, potentially opening new security vulnerabilities if not continuously monitored and updated \cite{chiejina2024system, aryal_moving_2023}.
}

Therefore, it is crucial to implement robust security measures that prevent, detect, and respond to such attacks targeting AI/ML components within O-RAN deployments \cite{groen2023implementing}.

Furthermore, a shift from centralized to distributed real-time applications is necessary to enhance security and efficiency. The decentralized nature of O-RAN complicates the security landscape, exposing AI models to various vulnerabilities, such as adversarial attacks, model poisoning, and data leakage. These risks are exacerbated by the multi-vendor environment inherent to O-RAN, which can lead to inconsistent security implementations across the network. 
{\color{black} This multi-vendor ecosystem introduces additional supply chain risks, where third-party xApps and rApps from various vendors may contain hidden vulnerabilities or backdoors, either through compromised development pipelines or intentional malicious code injection \cite{chiejina2024system}. The lack of transparent model verification and validation standards across vendors complicates the ability to audit AI components for integrity, further expanding the attack surface \cite{aryal_moving_2023}.
}
To address these challenges, FRL has been proposed as a viable solution, enabling collaborative model training while maintaining data privacy by keeping sensitive information localized on devices. This approach not only mitigates the risks associated with data transmission but also allows for enhanced resilience against attacks targeting AI models. Effective strategies, including the adoption of Distributed Ledger Technologies (DLTs), may provide enhancements in securing AI model operations by ensuring data integrity, facilitating secure identity management, and establishing automated collaboration protocols among diverse stakeholders in the O-RAN architecture \cite{aryal_moving_2023}. 

{\color{black} DLTs can provide immutable audit trails of model training processes and create cryptographic verification mechanisms for federated learning parameters, though they require careful design to maintain the real-time performance requirements of O-RAN systems \cite{aryal_moving_2023}.
}

{\color{black}Attackers can exploit adversarial inputs to manipulate the behavior of ML models, leading to inaccurate predictions and resource misallocation. For instance, malicious users may employ evasion attacks to fool the ML systems into making erroneous decisions, such as frequent and unnecessary handovers between cells, resulting in resource exhaustion and degraded service quality. 

These evasion attacks represent only the most visible threat from adversarial ML. A comprehensive threat model must also encompass model inversion attacks where adversaries reconstruct training data characteristics from model outputs, potential information leakage through model predictions that reveal proprietary network optimization strategies, and adversarial transferability where attacks designed against one model transfer to others with high success rates \cite{chiejina2024system}. The open nature of O-RAN architecture amplifies these risks, as adversaries with network access can perform high-volume queries against deployed ML models to extract parameters or functionality \cite{chiejina2024system}.  To mitigate these risks, it is crucial to implement robust defense mechanisms such as adversarial training, which incorporates adversarial examples into the training datasets, thereby enhancing the models' resilience to malicious inputs \cite{chen2023brief}. 

However, adversarial training itself introduces complex tradeoffs: overfitting to known adversarial examples can reduce model robustness to novel attacks, and computational overhead may be prohibitive in resource-constrained RAN environments \cite{chiejina2024system}. Defense mechanisms must therefore be combined with complementary strategies, including input validation, anomaly detection during inference, and robust model architectures specifically designed for the O-RAN domain \cite{groen2023implementing}.

These issues can also be addressed using eXplainable AI (XAI) techniques, which help human operators understand and manage AI decisions, thereby reducing the human-to-machine barrier and improving trust in AI systems \cite{brik_explainable_2025}. 
XAI becomes particularly critical for security in O-RAN contexts, as it enables operators to identify anomalous model behaviors that may indicate successful attacks or model drift, validate that models behave according to intended specifications, and detect potential vulnerabilities such as biased decision-making that could be exploited by adversaries \cite{brik_explainable_2025}. Additionally, \cite{moore2023toward} presented the idea of secure slicing using SliceX, an xApp designed to protect RAN resources and guarantee that performance standards are fulfilled even when malicious activity is present, proving its usefulness in actual situations.  In \cite{fiandrino2023explora}, the EXPLORA framework is proposed to enhance the transparency of DRL-based control solutions in the O-RAN ecosystem by making their decision-making process more understandable. EXPLORA generates network-oriented explanations using an attributed graph that links the actions executed by a DRL agent to the input state space. Each node in the graph contains relevant attributes that provide insight into why specific decisions were made, helping operators interpret, debug, and optimize AI-driven network management. As the framework EXPLORA has shown, this is very important for the understanding and mitigation of security risks of DRL-based control solutions within O-RAN. By providing clear insights into how decisions are made, EXPLORA helps identify potential vulnerabilities, such as biased or unsafe actions taken by the DRL agent. This transparency allows operators to detect and address anomalous behaviors, prevent adversarial attacks, and ensure that AI-driven controls do not compromise network security.}

To conclude, resolving AI security issues in O-RAN systems requires a multifaceted approach that includes strong adversarial defenses, transparent AI practices, distributed AI strategies, and thorough security evaluations and standardizations. Although AI and ML hold significant promise for enhancing O-RAN capabilities, their implementation must be carefully managed with robust security measures to mitigate risks and ensure the network's integrity and privacy. Table \ref{tab:security_challenges} summarizes the various security challenges faced in O-RAN and outlines corresponding ML solutions that can be employed to mitigate these risks, highlighting the critical role of ML in enhancing the security framework of O-RAN architectures.

\subsubsection{Case Study: ML-Driven DDoS Detection in O-RAN}
{\color{black}
To demonstrate the integration of AI/ML for real-world O-RAN security, we simulated a distributed denial-of-service (DDoS) attack scenario following the methodology presented in \cite{soleymani2024ddos}. The proposed framework deploys specialized ML-based applications within the O-RAN architecture, namely, a distributed application (dApp), an xApp for suspicious UE behavior detection (xApp-U), and an xApp for service usage monitoring (xApp-S). The system emphasizes real-time, localized monitoring within the RAN for fast detection, complemented by aggregated and context-rich analysis at the near-RT RIC layer.

The simulation utilized a dataset of multi-cell O-RAN traffic containing throughput, signal quality, and service usage data from multiple UEs across different gNBs \cite{xtephv3622,siriwardhana_descriptor_2025}. Various ML algorithms—including Random Forest (RF), Multilayer Perceptron (MLP), K-Nearest Neighbor (KNN), Decision Tree (DT), XGBoost (XGB), Support Vector Classifier (SVC), AdaBoost, Quadratic Discriminant Analysis (QDA), and Isolation Forest (IF) were trained and evaluated on standard metrics such as accuracy and F1-score.  Accuracy measures the overall proportion of correctly classified samples, while the F1-score provides a balanced assessment of precision and recall, particularly useful in the presence of class imbalance.

As shown in Fig. \ref{fig:pict_accuracy} and Fig. \ref{fig:pict_f1}, models such as RF, MLP, and KNN achieved the highest detection accuracies (above 99.9\%) and F1-scores close to 1.0, indicating near-perfect classification of malicious traffic patterns. Ensemble models like RF and XGBoost exhibited strong generalization and computational efficiency, making them suitable for near-real-time detection in the RIC. In contrast, simpler or unsupervised models such as QDA and IF showed lower reliability (accuracy of 80.48\% and 62.39\%, respectively), highlighting the importance of selecting appropriate algorithms based on deployment context.

Overall, this case study validates the effectiveness of ML-based detection frameworks in safeguarding O-RAN against volumetric and behavior-based DDoS attacks. Beyond detection, the proposed architecture demonstrates how AI/ML-driven intelligence can be embedded into the O-RAN control loop to enable autonomous network protection. Ultimately, adaptive intelligence in O-RAN marks a step toward networks that secure and optimize themselves in real time.

\begin{figure}[h!]
\centering
\includegraphics[width=3.5in]{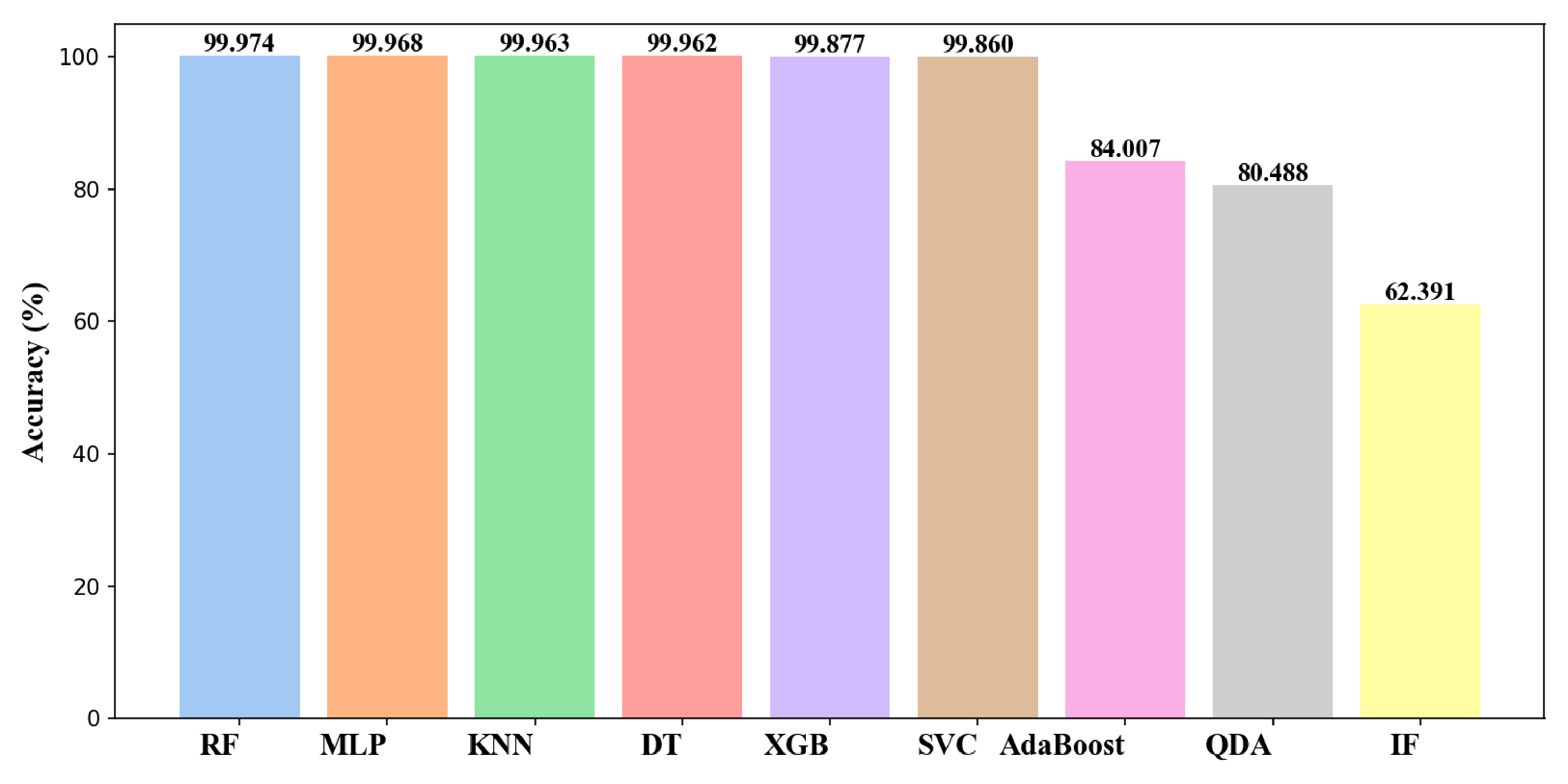}
\caption{{\color{black} Comparison of Accuracy for ML Algorithms}}
\label{fig:pict_accuracy}
\end{figure}

\begin{figure}[h!]
\centering
\includegraphics[width=3.5in]{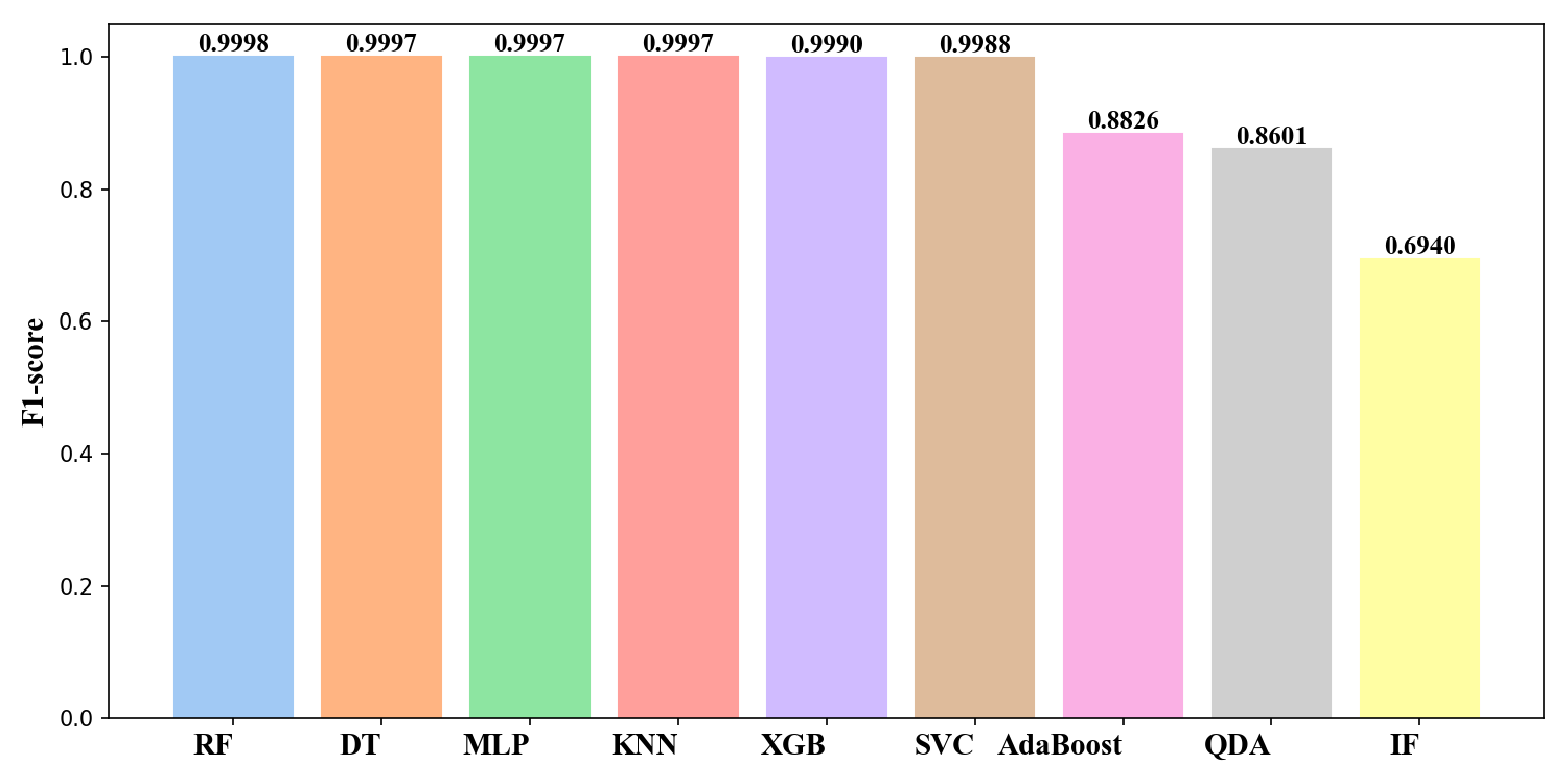}
\caption{{\color{black} Comparison of F1-score for ML Algorithms}}
\label{fig:pict_f1}
\end{figure}
}

\subsection{Lessons Learned}
\begin{itemize}
    \item {\color{black} \textbf{Spectrum Management:} Dynamic spectrum management in O-RAN environments becomes more efficient and feasible through the integration of third-party applications on the RICs to perform near-real-time and long-term optimization strategies. Various AI models are being explored depending on the specific strategy being proposed. RL and DRL (Q-learning, DQN, PPO, Actor/Critic) models stand out as a solution to dynamic spectrum sharing or channel selection across bands, as they enable SUs to learn optimal access policies that maximize spectrum utilization while respecting interference constraints and avoiding harmful interference to PUs. The advantage of RL/DRL is their ability to continuously learn optimal access and power control policies through interacting with the environment \cite{10182973,10646359,10794361}, efficiently adapting to dynamic channel conditions and traffic variations without requiring explicit system modeling. Nevertheless, other AI/ML techniques such as CNNs, RNNs, LSTMs, statistical learning approaches, and gradient-boosted trees are also useful to predict the users' traffic, manage the interference, sense the spectrum, ultimately increasing the efficiency of the spectrum utilization. AI/ML algorithms will remain at the core center of next-generation wireless networks, thanks to their ability of learning network demand patterns and dynamically allocate resources.

    \item \textbf{Resource Allocation:} While traditional, rule-based, and monolithic optimization methods have been effective in earlier network generations, they face limitations in addressing the increasingly dynamic, sliced, and disaggregated nature of modern networks that must support diverse users and services. Instead, AI/ML methods such as multi-agent DRL models, graph neural networks (GNNs), Bayesian optimization, clustering, and supervised classifiers are well-suited for multiobjective resource allocation. They leverage O-RAN’s near-RT and long-term telemetry, continuously interacting with the network and adapting resource management policies to real-time conditions and diverse QoS requirements. However, one should particularly be meticulous in designing the AI/ML models to achieve truly optimal allocation and a fair balance between the objectives, such as power consumption, latency, and QoS threshold compliance.
}

\item {\color{black} \textbf{Security:} The critical role of AI/ML in strengthening O-RAN security is highlighted through their ability to enable adversarial defenses, explainable frameworks such as EXPLORA, secure slicing, encryption, and intelligent threat detection. By integrating robust, explainable, and distributed AI solutions, O-RAN can protect open interfaces, data privacy, and supply chains while ensuring network integrity and resilience in multi-vendor, hyperconnected environments.

The analysis of different ML approaches applied in O-RAN highlights that no single method is universally optimal for addressing all security challenges. Instead, the choice of technique should depend on the nature of the threat, data availability, and operational constraints. UL techniques, such as clustering and autoencoders, are particularly well-suited for intrusion and anomaly detection in dynamic O-RAN environments, where labeled data may be scarce. SL methods, including support vector machines and deep neural networks, are more effective for known threat classification and predictive defense strategies but rely on large, well-curated datasets. RL enables adaptive and autonomous responses to evolving security threats, making it ideal for real-time attack mitigation, though it introduces concerns related to stability and explainability. DL architectures such as CNNs and LSTMs have shown strong performance in authentication and access control, especially for device fingerprinting, but they may be vulnerable to adversarial manipulation. FL, often combined with privacy-preserving mechanisms, is promising for protecting sensitive data during collaborative security training across multiple domains. Finally, robustness techniques like knowledge distillation and adversarial training strengthen ML models against evasion and poisoning attacks, although they require careful tuning. These insights underscore that effective O-RAN security will likely rely on hybrid ML strategies, integrating complementary strengths from multiple learning paradigms to build resilient, adaptive, and explainable defense mechanisms.
}
\end{itemize}

\begin{table*}[h!]
\centering
\caption{Security Challenges in O-RAN and Corresponding ML Solutions}
\label{tab:security_challenges}
\begin{tabular}{|p{5cm}|p{4.1cm}|p{4cm}|p{0.1cm}|}
\hline
\multicolumn{1}{|c|}{\textbf{Security Challenges}} & \multicolumn{1}{|c|}{\textbf{Description}} &	\multicolumn{1}{|c|}{\textbf{ML Solutions}} & \multicolumn{1}{c|}{\textbf{References}} \\
\hline  

\multicolumn{1}{|c|}{\textbf{\multirow{4}{*}{Network Architecture \& Interoperability}}} & Open, standardized interfaces increase attack surfaces, making O-RAN more vulnerable to security threats. &	SL for cell traffic prediction, DRL for energy-efficiency, and adversarial defense models. & \multicolumn{1}{c|}{\multirow{4}{*}{\cite{liyanage2023open,sapavath2023experimental,giannopoulos_supporting_2022}}} \\ \hline

\multicolumn{1}{|c|}{\textbf{\multirow{4}{*}{Ecosystem \& Vendor Security }}} & Multi-vendor environments increase risks of supply chain attacks, including tampering and supply chain poisoning. & UL for anomaly detection, predictive analytics for threat intelligence, and dynamic security mechanisms. & \multicolumn{1}{c|}{\multirow{4}{*}{\cite{yeboah-ofori_cyber_2021,chen2023detecting,afaq2021machine}}} \\ \hline

\multicolumn{1}{|c|}{\textbf{\multirow{4}{*}{Data Protection \& Privacy}}} & Ensuring privacy and protection of sensitive data within a hyperconnected, open network, especially in 6G networks. & AI/ML-based encryption methods, DL models for secure data handling, and anomaly detection for unauthorized access. & \multicolumn{1}{c|}{\multirow{4}{*}{\cite{gajjar2024preserving,9881863,rahman2021network,puppo2024extraction}}} \\ \hline

\multicolumn{1}{|c|}{\textbf{\multirow{4}{*}{AI Trust \& Reliability}}} & Vulnerabilities of AI models to adversarial attacks that can degrade performance, necessitating robust defense strategies. & Adversarial defense mechanisms, Explainable AI (XAI) for transparency, Federated Learning for secure, distributed AI. & \multicolumn{1}{c|}{\multirow{4}{*}{\cite{chiejina2024system,brik_explainable_2025,moore2023toward,kouchaki2023openai}}} \\ \hline

\multicolumn{1}{|c|}{\textbf{\multirow{4}{*}{Network Automation \& Security}}} & Managing the complexity of multivendor and interoperable solutions while maintaining security in an open and dynamic network. & ML-driven automation for threat detection, RL for adaptive security policies, ML-enhanced access control. & \multicolumn{1}{c|}{\multirow{4}{*}{\cite{anurag2024robotic,machhindra2023enhancing,hamdan_recent_2023}}} \\ \hline

\multicolumn{1}{|c|}{\textbf{\multirow{4}{*}{Open-Set Device Identification}}} & Identifying unauthorized devices and preventing unauthorized access in an open and flexible network architecture. & CNN+LSTM models for RF signal pattern recognition, open-set detection approaches for device identification. & \multicolumn{1}{c|}{\multirow{4}{*}{\cite{puppo2024extraction,chiejina2024system}}} \\ \hline

\multicolumn{1}{|c|}{\textbf{\multirow{4}{*}{Scalability and Performance}}} & Ensuring that ML-based security solutions scale effectively and perform efficiently as networks expand and handle more data. & Optimization of ML algorithms for large-scale data processing, cross-network ML model deployment, and evaluation. & \multicolumn{1}{c|}{\multirow{4}{*}{\cite{afaq2021machine,mittal2023ai}}} \\ \hline

\end{tabular}
\end{table*}

{\color{black}
\section{Paving the Path Forward: Future Directions for ML in O-RAN}}
{\color{black}
The role of ML in supporting and advancing O-RAN technology, as mentioned in the previous sections, is becoming increasingly essential to meet the challenges of future network implementations that are more dynamic and complex. With the increasing need for more efficient networks, further research is urgently needed to address the emerging constraints and study the potential of ML implementation in  O-RAN. The open nature of the O-RAN, a key aspect that drives interoperability and reduces costs, also makes it vulnerable to potential privacy and trust issues. Moreover, when combined with the complex environment in which O-RAN operates, it may lead to conflicting actions that require solutions. Therefore, this section presents promising future research directions for applying ML in O-RAN. It focuses on areas such as conflict mitigation in multi-component systems, mmWave \cite{keyela2022discrete}, and Terahertz integration, scalability, and performance optimization, ultra-massive MIMO for coverage enhancement, and improving efficiency through mobile edge computing.

\subsection{Towards Conflict Mitigation in Multi-Component O-RAN Systems}
Due to their considerable complexity, conflicting actions will likely arise in O-RAN environments. A complex environment, characterized by the participation of many entities in a decentralized decision-making system, may suffer from impaired coordination and overall network efficiency due to possible conflicts of actions arising from decisions made by various entities. When multiple logical controllers in an O-RAN make conflicting or disruptive decisions, conflicts may occur, hindering the overall performance and efficiency of the network. Conflicts may arise between xApps \cite{adamczyk_detection_2023, brach_del_prever_pacifista_2025, erdol_xapp_2026, shami2025ran}, conflicts of intents and policies \cite{de_oliveira_towards_2024}, and resource conflicts \cite{lozano_airic_2024}. As a result, conflict resolution in a decentralized O-RAN environment is challenging.

Conflict mitigation entails detecting, preventing, and resolving issues that may arise between decisions made by various entities.  While there has been some notable research into conflict mitigation, it is still in-depth and very limited in its utilization of AI/ML. Therefore, it is encouraged that future research focuses on exploring more advanced and adaptive detection techniques, which can utilize AI/ML for real-time conflict prediction and mitigation. For example, the combined implementation of FL and Multi-Agent RL (MARL) is a promising solution, as conflict mitigation requires a dynamic mechanism that can adjust resources, coordinate policies, and synchronize application operations in real time. Both have advantages and disadvantages that can complement each other to prevent or mitigate conflicts in distributed systems such as O-RAN. The FL architecture allows local operations and policies that enhance data privacy \cite{erdol_federated_2022}; however, each local model cannot collaborate and communicate, making it less effective in resource distribution and conflict mitigation because all communication must go through the center. On the other hand, MARL architecture is capable of efficiently communicating and collaborating with fellow agents \cite{Rezazadeh_Multi_2023}, as a result, their collaboration will be very effective in preventing and mitigating the conflict of policies and resource allocation in O-RAN.

\subsection{Advancing Millimeter-Wave and Terahertz Integration} 
O-RAN, with its open nature, aims to create more flexible, scalable, and multi-vendor networks. However, the highly reliable low-frequency spectrum (Sub-6 GHz) is increasingly competitive in use due to the physical limitations of its spectrum allocation. While O-RAN adds to the diversity of devices and providers by keeping spectrum coordination and control efficient and adaptive, its dependence on efficient and flexible spectrum allocation complicates its expansion \cite{sharma_overview_2025}. The increasing demand for high-speed connectivity and more efficient communications is making spectrum management a key challenge that requires more serious attention. 

The mmWave (30-300 GHz) and THz (0.1-10 THz) \cite{khan_ai-ran_2024} spectrum can be a solution to overcome the limitations of the Sub-6 GHz spectrum, as they enable high-capacity data transfer, thereby reducing spectrum congestion. Recently, mmWave is the primary communication solution in 5G and B5G, where its deployment already has official standards and is supported by existing devices \cite{kumar_review_2024}.  It has better coverage but still has limited bandwidth, and its coverage range is limited due to high path loss and sensitivity to blockage.  In addition, even though THz is still being researched as a key technology for 6G, it has an extensive bandwidth and a very high data rate \cite{kumar_review_2024}. However, due to its mmWave and THz characteristics, it is susceptible to channel changes, obstacles, and atmospheric absorption. Reconfigurable Intelligent Surfaces (RIS) are programmable surface structures that control the reflection of electromagnetic (EM) waves, which have the potential to overcome these limitations \cite{10921906}. Dynamically, RIS can reflect and modify electromagnetic waves to enhance signal strength and range, which enables more reliable and flexible mmWave and THz communication and installations. The complex configuration of RIS makes it challenging to control and optimize, rendering its performance in supporting mmWave and THz communication in O-RAN ineffective. Therefore, ML becomes a critical component in enabling the use of RIS. ML can be used to optimize the RIS reflection phase efficiently, and even the combination of RIS with ML can overcome signaling overhead and fast channel setup, especially in dense environments typical of mmWave/THz \cite{qurratulain_khan_machine_2023}. Thus, RIS integrated with ML is a fundamental technology that enables the efficiency and coverage of high-frequency communication systems. Therefore, further research in this area is essential, as RIS-assisted integration of mmWave and THz in O-RAN could be the key to realizing the full potential of 6G networks.}

{\color{black}
\subsection{Scalability and Performance Optimization in Large-Scale O-RAN}
ML techniques play a crucial role in various aspects of O-RAN, including security, traffic optimization, and resource management, by enabling real-time decision-making. However, the growing scale of these networks presents challenges related to the efficient processing of large data volumes, the optimization of computational resources, and the deployment of ML models that can operate seamlessly across different network layers. Therefore, as O-RAN networks continue to expand, it is essential to ensure that ML-based solutions can scale effectively to accommodate increasing traffic volumes and network complexity. 

To address scalability challenges, federated learning offers a promising solution by enabling decentralized training across distributed edge nodes \cite{tak2020federated}. This approach reduces the need for large-scale data transfers, thereby enhancing privacy and minimizing latency while allowing ML models to adapt to dynamic network conditions \cite{singh_communication_2024}. This decentralized approach supports scalability by allowing parallel model training at the edge, minimizing latency and central bottlenecks, and enabling efficient adaptation to growing network size and complexity in large-scale O-RAN deployments. The integration of model compression techniques, such as knowledge distillation and quantization, can also enhance efficiency by reducing the computational burden of ML models without compromising performance. Knowledge distillation enables a smaller model to learn from a larger, more complex model, retaining its accuracy while requiring fewer resources. Similarly, quantization reduces the precision of numerical computations, decreasing memory usage and accelerating processing \cite{dantas2024comprehensive}. These methods support scalability by enabling the deployment of lightweight models across many edge nodes, ensuring efficient performance as O-RAN networks grow in size and complexity. Although federated learning and model compression show promise in simulations, there is limited work validating their effectiveness in practical deployments, highlighting a key gap in current research. Hence, future research should evaluate these approaches under real-world, high-load O-RAN conditions. 

\subsection{Exploring Ultra-Massive MIMO for O-RAN Coverage Enhancement}
O-RANs are currently challenged by issues related to coverage and spectrum efficiency, particularly as the demand for high data rates and reliable connectivity continues to escalate. While traditional MIMO (Multiple Input Multiple Output) technologies have provided substantial enhancements to system performance, Ultra-Massive MIMO represents an advanced evolution of this technology that utilizes a far larger number of antennas at both the transmitter and receiver. This innovation significantly improves coverage and diversity. Specifically, Ultra-Massive MIMO harnesses hundreds or even thousands of antennas to simultaneously serve multiple users, enabling optimal spatial multiplexing and advanced interference mitigation \cite{huo2023technology}. As a result, this technology enhances user experience through improved throughput and reduced latency, addressing the stringent demands of 5G and future wireless systems \cite{nie2019intelligent}.

The integration of Ultra-Massive MIMO into the O-RAN architecture can amplify these benefits further. By leveraging ML, O-RAN can adaptively optimize antenna configurations based on real-time channel conditions and user demands. ML algorithms can analyze extensive datasets to predict user locations, develop optimal beamforming strategies, and dynamically allocate resources based on expected traffic patterns. This approach not only maximizes the performance of Ultra-Massive MIMO systems but also ensures that the network remains robust and responsive to user needs. Furthermore, given its significant potential to enhance O-RAN performance, further research into Ultra-Massive MIMO is both timely and necessary. Researchers are encouraged to explore and implement ML solutions that facilitate real-time optimization of Ultra-Massive MIMO in O-RAN environments. Such collaborations will pave the way for the development of more resilient and efficient wireless networks, contributing meaningfully to the evolution of next-generation communication standards.

\subsection{Efficient Integration of MEC and O-RAN}
As O-RAN networks expand, they increasingly encounter challenges related to latency and bandwidth, particularly for applications requiring real-time processing and high data rates, such as augmented reality and IoT services. Mobile Edge Computing (MEC) emerges as a significant solution by decentralizing computation resources closer to end-users, thereby addressing these pressing issues. By positioning computing capabilities at the network edge rather than solely depending on centralized cloud servers, MEC effectively reduces latency. This configuration minimizes the distance that data must travel, subsequently enhancing response times and enabling immediate data processing—crucial for applications that demand low latency. Furthermore, MEC alleviates network congestion by offloading compute-intensive tasks from the core network, thereby maximizing resource utilization and improving overall user experiences \cite{ali_mobile_2018}.

Integrating MEC within the O-RAN framework offers a promising approach for addressing these challenges while enhancing network efficiency \cite{saguna2016using}. When combined with ML, operators can establish localized compute resources that work in conjunction with advanced algorithms for dynamic resource management. For instance, ML can predict traffic loads and optimize resource allocation across edge nodes, significantly minimizing potential bottlenecks. Moreover, the intelligent caching of frequently accessed data can be implemented through ML, ensuring that necessary information is stored closer to users and enhancing network speed and performance. The integration of MEC, O-RAN, and ML represents an inviting area for future research and development. Scholars and researchers are urged to focus on crafting innovative ML models that strengthen the integration of MEC and O-RAN. This focus will help address emerging challenges in the wireless landscape while simultaneously improving service quality and operational efficiency. 

{\color{black}
\subsection{Leveraging Digital Twin Technology to Achieve URLLC in O-RAN}
The integration of Digital Twin (DT) \cite{tao_digital_2023} technology within the O-RAN architecture holds great promise for achieving the stringent URLLC KPIs required by next-generation wireless services. By creating accurate, real-time virtual representations of physical network components, DTs enable continuous network monitoring, predictive analytics, and dynamic resource allocation, ultimately improving reliability and reducing latency. These capabilities allow the network to proactively adapt to changing traffic patterns, anticipate failures, and optimize resource utilization with minimal disruption to ongoing services \cite{masaracchia2025toward}.

Beyond its technical capabilities, the concept of the DT aligns closely with the fundamental principles of the O-RAN Alliance, i.e., openness, intelligence, and autonomy. Both O-RAN and DT to are driving the evolution of next-generation RANs toward more flexible, adaptive, and self-optimizing architectures. DT and O-RAN form two synergistic paradigms that together can facilitate the development of a smart, resilient, and transparent 6G RAN capable of supporting emerging applications and services \cite{masaracchia2025digital}.

However, turning this potential into reality comes with important challenges, especially when it comes to keeping physical systems and their digital counterparts in sync in real time. As more sensors are deployed in advanced 6G scenarios, the amount of data sent from IoT devices to edge and cloud servers grows rapidly. This surge in traffic can put significant pressure on network resources, making it harder to maintain the ultra-low latency and high reliability that URLLC demands. Ensuring precise and continuous synchronization is therefore essential to keep the digital representation accurate and fully aligned with the physical world \cite{masaracchia2023digital}.

Looking ahead, future research should focus on designing scalable DT orchestration frameworks, edge-intelligent synchronization mechanisms, and lightweight predictive models that minimize processing delays while maintaining high fidelity. Integrating DTs with AI-driven control loops can enable adaptive decision-making for real-time resource optimization, proactive fault management, and enhanced situational awareness. These advancements will be key enablers of URLLC in next-generation O-RAN deployments, supporting emerging applications such as remote surgery, industrial automation, and autonomous systems.

\subsection{Lessons Learned}
    \begin{itemize}
    {\color{black}
    \item Understanding the root causes of conflicts is critical for effective conflict-mitigation strategies in O-RAN. Such conflicts can arise not only from differing xApp tasks and objectives but also from intent and policy discrepancies, as well as competition for shared resources. Mitigating these conflicts requires secure, adaptive, and flexible mechanisms capable of distributing policies across decentralized systems, managing resources efficiently, and maintaining real-time synchronization. Privacy-preserving approaches—such as performing operations locally without exposing sensitive data—can be highly beneficial when combined with coordination mechanisms that ensure essential updates are shared across agents. However, enforcing privacy without any means of synchronizing key updates can impede conflict resolution, highlighting the need for a balanced approach between privacy and coordination.

    \item Innovations in mmWave (30–300 GHz) and THz (0.1–10 THz) spectrum technologies are highly compatible with O-RAN’s stringent requirements for ultra-high-speed, low-latency, and energy-efficient communications. Leveraging these high-frequency bands presents unique challenges, including severe propagation loss, sensitivity to blockage, and limited coverage, which necessitate advanced beamforming, intelligent resource allocation, and dynamic spectrum management. Future O-RAN spectrum management solutions must account for these distinct characteristics while enabling seamless coordination and aggregation across mmWave and THz bands to meet diverse and demanding user requirements. Successfully integrating these bands requires not only innovations across all layers of system design, from PHY/MAC to network orchestration, but also a comprehensive understanding of how mmWave and THz can complement each other to achieve a unified, flexible, and efficient O-RAN architecture.}

    \item The application of ML in O-RAN not only brings advanced capabilities that can make networks more intelligent, more adaptive, and more efficient, but also raises new issues around real-time coordination and scalability. Through this study, it has become clear that this requires an approach that can help reduce latency and enable real-time optimization, such as integrating ML with Ultra-Massive MIMO and MEC.

    \item The development of DT technology utilization increasingly demonstrates how ML can improve prediction and control in complex systems. These developments indicate that the success of ML in O-RAN depends not only on technical advances, collaboration, and continuous testing in real-world environments, but also on monitoring, simulation, and configuration to build complex, intelligent, reliable, and future-ready network systems.
    
    \iffalse
    \item Through this study, it became clear that applying ML in O-RAN is both promising and challenging. ML brings powerful capabilities that can make networks smarter, more adaptive, and more efficient, but it also introduces new issues around coordination, scalability, and trust. We learned that conflicts between different components in an open and distributed system like O-RAN are unavoidable, and effective solutions must focus on smarter, cooperative decision-making. Techniques such as federated learning and multi-agent reinforcement learning can help balance local autonomy with global coordination. We also found that ML plays a vital role in enabling advanced technologies such as mmWave, Terahertz, and Reconfigurable Intelligent Surfaces, which are key to improving coverage and reliability in high-frequency networks. Similarly, integrating ML with Ultra-Massive MIMO and Mobile Edge Computing helps reduce latency and make real-time optimization possible. The emerging use of Digital Twin technology further shows how ML can enhance prediction and control in complex systems. Overall, the main lesson is that the success of ML in O-RAN depends on more than technical progress. It requires collaboration, transparency, and continuous testing in real-world environments to build intelligent, reliable, and future-ready networks.
    \fi
    \end{itemize}
    }
    }
    
{\color{black}
\section{Conclusions}
The rapid growth of user demands places significant pressure on O-RAN to deliver seamless, high-performance connectivity, marking a transformative phase in the telecommunications industry. While O-RAN’s openness and integrated intelligence offer substantial benefits, they also introduce new challenges that require careful management and innovative solutions. This survey provides a comprehensive examination of AI/ML implementations within O-RAN, evaluating both the progress achieved and the outstanding challenges in critical areas such as spectrum management, resource allocation, and security. Advances in AI/ML have enabled effective, adaptive solutions across these domains, with each ML paradigm contributing according to its unique characteristics and strengths. By leveraging these capabilities, ML-driven approaches can dynamically optimize network performance, improve decision-making, and uphold stringent quality-of-service standards. Additionally, this survey outlines future research directions that remain essential for the continued evolution of intelligent O-RAN systems. Overall, our analysis underscores that AI/ML has become an integral component of O-RAN, guiding its development along a strategic, adaptive, and technology-driven trajectory.}

% or


\ifCLASSOPTIONcaptionsoff
  \newpage
\fi

\end{document}